\documentclass[twocolumn]{aastex631}

\definecolor{burntorange}{rgb}{0.8, 0.33, 0.0}

\newcommand\lsim{\mathrel{\rlap{\lower4pt\hbox{\hskip1pt$\sim$}}
    \raise1pt\hbox{$<$}}}
\newcommand\gsim{\mathrel{\rlap{\lower4pt\hbox{\hskip1pt$\sim$}}
    \raise1pt\hbox{$>$}}}

\usepackage{amsmath}

\begin{document}

\title[Gaia BH1 Triple]{A Triple Scenario for the Formation of Wide Black Hole Binaries Such As Gaia BH1}

\correspondingauthor{Aleksey Generozov}
\email{aleksey.generozov@gmail.com, hperets@physics.technion.ac.il}

\author[0000-0001-9261-0989]{A. Generozov}
\affiliation{Technion - Israel Institute of Technology\\
Haifa, 3200003, Israel}

\author[0000-0002-5004-199X]{H. B. Perets}
\affiliation{Technion - Israel Institute of Technology\\
Haifa, 3200003, Israel}

\submitjournal{ApJ}

\begin{abstract}
Recently, several non-interacting black hole-stellar binaries have been identified in Gaia data. For example, Gaia BH1, where a Sun-like star is in a moderate eccentricity (e=0.44), 185-day orbit around a black hole. This orbit is difficult to explain through binary evolution. The present-day separation suggests the progenitor system would have undergone an episode of common envelope evolution, but a common envelope should shrink the period below the observed one. Since the majority of massive stars form in higher multiplicity systems, a triple evolution scenario is more likely for the progenitors of BH binaries. Here we show that such systems can indeed be more easily explained via evolution in hierarchical triple systems. von Zeipel--Lidov--Kozai oscillations or instabilities can delay the onset of the common envelope phase in the inner binary of the triple, so that the black hole progenitor and low-mass star are more widely separated when it begins, leading to the formation of wider binaries. There are also systems with similar periods but larger eccentricities, where the BH progenitor is a merger product of the inner binary in the triple. 
Such mergers lead to a more top-heavy black hole mass function.
\end{abstract}



\section{Introduction}
Approximately one out of every thousand stars will end their lives as a black hole (BH). This means a galaxy like the Milky Way 
should contain of order $10^8$ BHs. However, only of order a hundred BH systems are known in our Galaxy, predominantly from studies 
of X-ray binaries \citep{corral-santana+2016,fortin+2023}, that can only identify close, interacting systems.

Microlensing \citep{lam+2022,mroz+2022,sahu+2022}, spectroscopy \citep{shenar+2022}, and astrometry \citep{andrews+2022,shahaf+2023} can be used to identify isolated BHs or non-interacting binaries.  Such systems probe high mass binary evolution over a wider range of parameter space than X-ray binaries, providing important constraints on the BH natal kick distribution and common envelope physics. 

Gaia DR3 contains a few$\times 10^5$ binary orbital solutions \citep{gaia+2023a,gaia+2023b}, prompting searches for astrometric BH binaries \citep{andrews+2022,shahaf+2023,chakrabarti+2023,elbadry+2023,elbadry+2023bh2}.
To date, two unambiguous BH-stellar binaries have been identified: Gaia BH1 and Gaia BH2. Gaia BH1 contains a G-type main sequence star in a 185.5-day orbit around a $9.32^{+0.22}_{-0.21}$ $M_{\odot}$ BH. The star's orbit is moderately eccentric with $e=0.44$ \citep{chakrabarti+2023,elbadry+2023}. Gaia BH2 contains a red giant in a 1276.7 day orbit around a $8.94 \pm 0.34$ $M_{\odot}$ BH. This star's orbit is also moderately eccentric with $e=0.52$ \citep{elbadry+2023bh2}. The uncertainties on the orbital periods and eccentricities are much less than 1\%.

These systems pose a challenge to isolated binary evolution models. 
On the one hand, the present-day separations suggest that the progenitor system would have undergone an episode of common envelope evolution. 
On the other hand, a common envelope would shrink the binary period below the observed one. 
The tension can be resolved by invoking an unusually large common envelope efficiency ($\alpha \approx5$ for Gaia BH1; \citealp{elbadry+2023}), 
or a very massive progenitor for the BH that avoids a red giant phase (and hence the common envelope) entirely \citep{elbadry+2023bh2}.

In any case, an isolated binary is not the most likely initial configuration for the progenitor system. At least 55\% of O star primaries have two or more companions with mass ratio, $q\geq 0.1$ \citep{moe&distefano2017}. Stellar evolution in triples could give rise to a variety of novel evolutionary channels and produce different types of binaries and merger products \cite[e.g.][]{perets&fabrycky2009,perets&kratter2012,Ant+14,Mic+14,naoz+2016,Ant+17,Ros+19,toonen+2020,Ste+22,hamers+2022,toonen+2022,kummer+2023,shariat+2023}.
Motivated by this observation we check whether triple evolution can reproduce the Gaia BH binaries. We find that triples can reproduce the observed period and eccentricity of Gaia BH1, even for normal common envelope efficiencies ($\alpha=1$).

The remainder of this paper is organized as follows. In \S~\ref{sec:methods} we discuss our initial conditions and procedure for evolving triples. 
In \S~\ref{sec:results} we present the results from our triple population synthesis and compare them to the results of binary population synthesis. In \S~\ref{sec:rates} we discuss formation rates. In \S~\ref{sec:alt} we discuss alternative formation scenarios. We summarize in \S~\ref{sec:conc}.

\section{Methods}
\label{sec:methods}
\subsection{Initial conditions}
We use the empirical distributions for multiple properties from \citet{moe&distefano2017} to initialize triples, with the modifications described below.

\begin{figure}
    \includegraphics[width=\columnwidth]{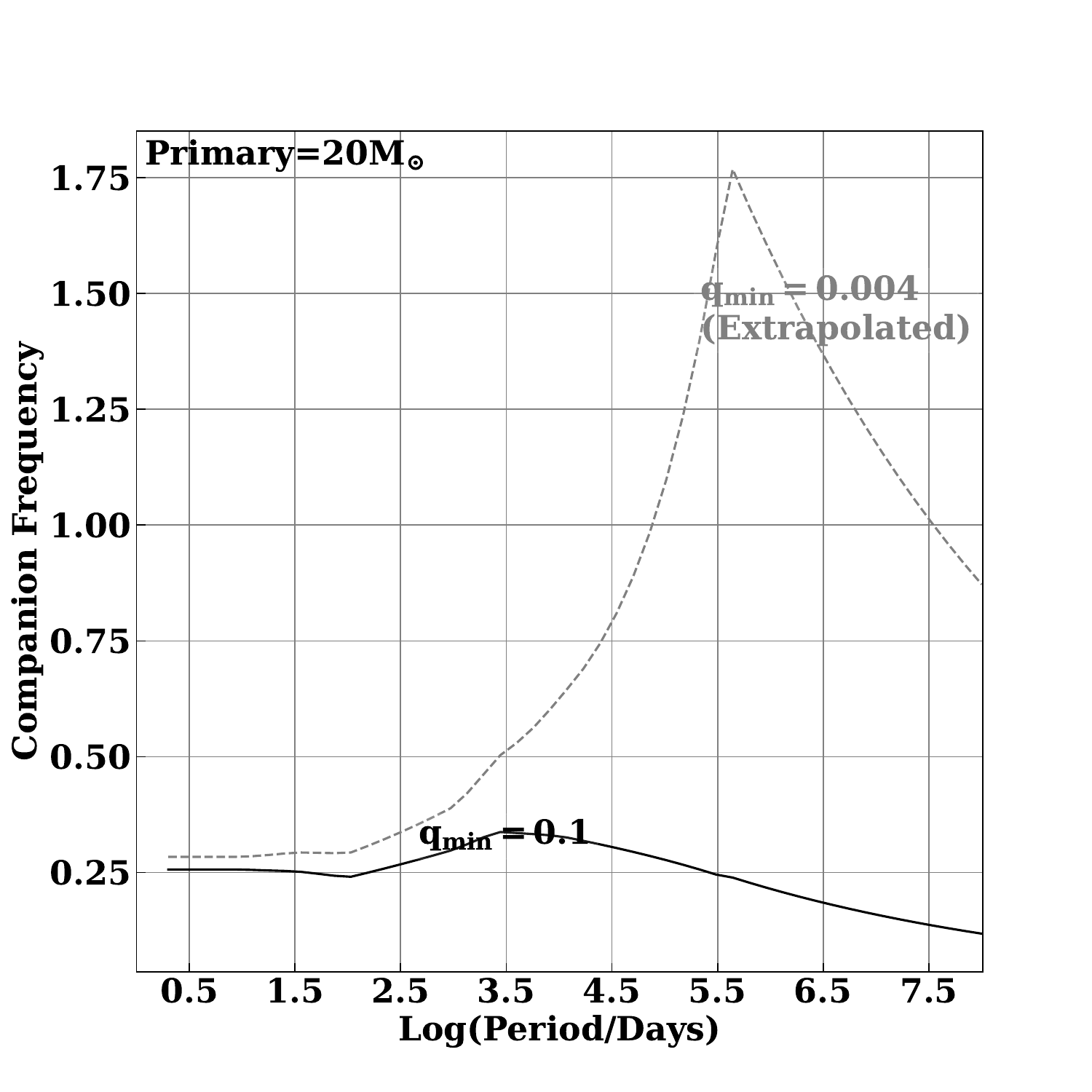}
    \figcaption{\label{fig:extrap} The solid, black line is the companion frequency (per decade orbital period) from \citet{moe&distefano2017} for a $20 M_{\odot}$ primary. The dashed, grey line shows the companion frequency, extrapolated to the brown dwarf limit.}
\end{figure}

The progenitor of the BH in Gaia BH1 would be $\gsim 20 M_{\odot}$, while its companion is 0.93 $M_{\odot}$.  Thus, the initial mass ratio is $\lsim 0.05$. The distribution of binary properties for such low mass ratios is unconstrained by \citet{moe&distefano2017}. Here, we extrapolate the broken power-law mass ratio distribution from their work down to the brown dwarf boundary. This extrapolation affects the distribution of companion periods, as shown in Figure~\ref{fig:extrap}.

To generate triples, we follow the procedure below:

\begin{enumerate}
    \item We draw a primary mass between 18 and 150 $M_{\odot}$ from an $m^{-2.3}$ distribution (corresponding to either a Kroupa or Salpeter mass function). 
    \item Then we generate two companion periods for this primary mass from its companion frequency distribution (modified by the extrapolation to lower mass ratios; see Figure~\ref{fig:extrap}), with the following steps
    \begin{enumerate}
        \item We first generate logarithmically spaced bins in period. The probability of a companion in each bin is the companion frequency multiplied by the bin width.
        \item We generate a random number between 0 and 1 for each bin (starting at the shortest period). If this number is less than the companion probability we generate a companion in this bin (with the period drawn from a log-uniform distribution). 
        \item The companions' eccentricities follow the distribution in \citet{moe&distefano2017}.\footnote{We include a turnover at 80\% of the maximum eccentricity, $1-\mathrm{\left(\frac{Period}{2 days}\right)}^{-2/3}$, to ensure the eccentricity distribution is continuous.} The mass ratio is drawn from the extrapolated mass ratio distribution. 
         \item We continue until we have a triple. For simplicity, we do not allow for quadruples or higher-order multiples in this study. 
    \end{enumerate}
\end{enumerate}

To form a Gaia BH-like binary, the progenitor system must contain a low-mass star. Thus, we require either the secondary or tertiary star to be between 0.5 and 2 $M_{\odot}$.
Finally, we only consider triples that satisfy the \citep{mardling&aarseth2001} stability criterion.

\begin{figure*}
    \plotone{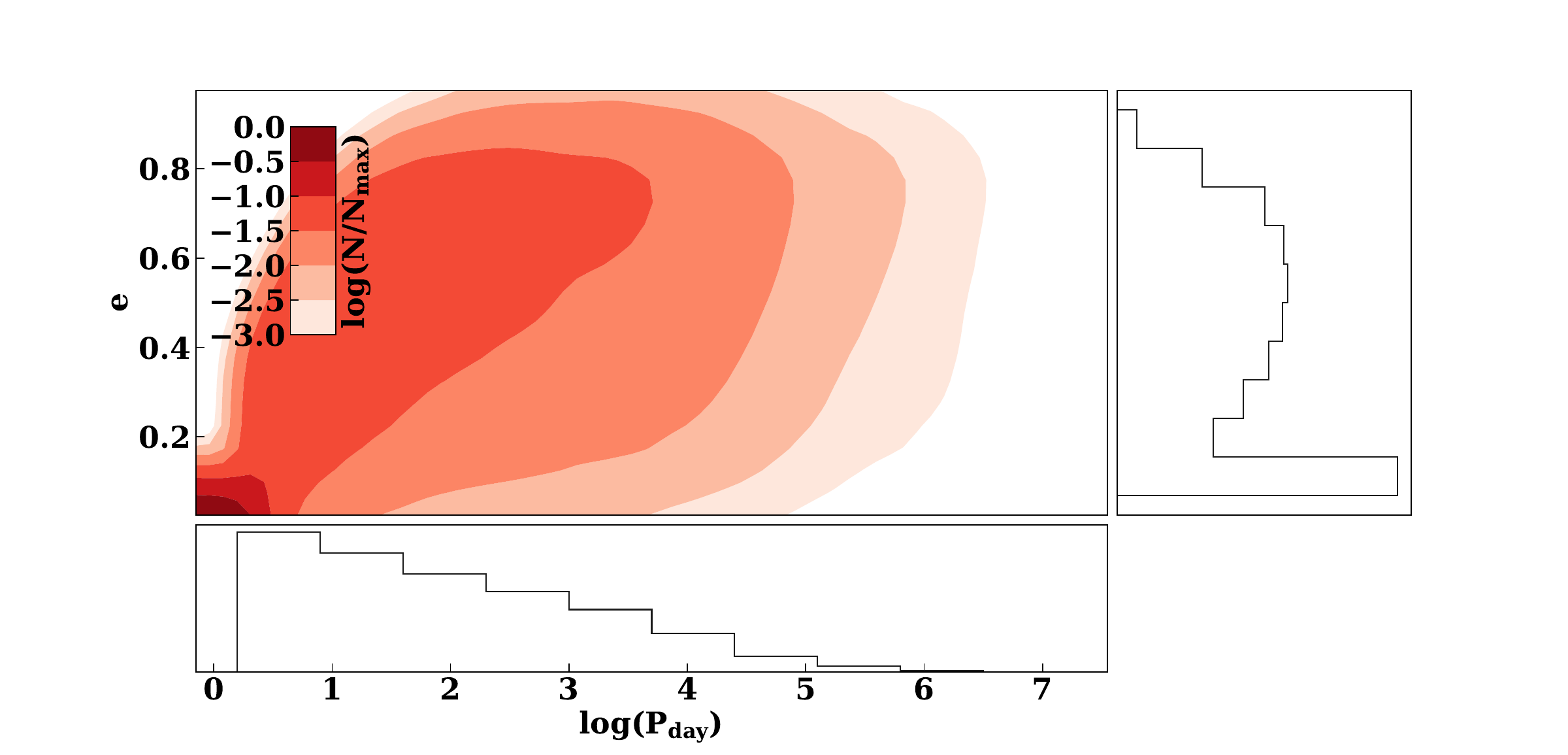}
    \plotone{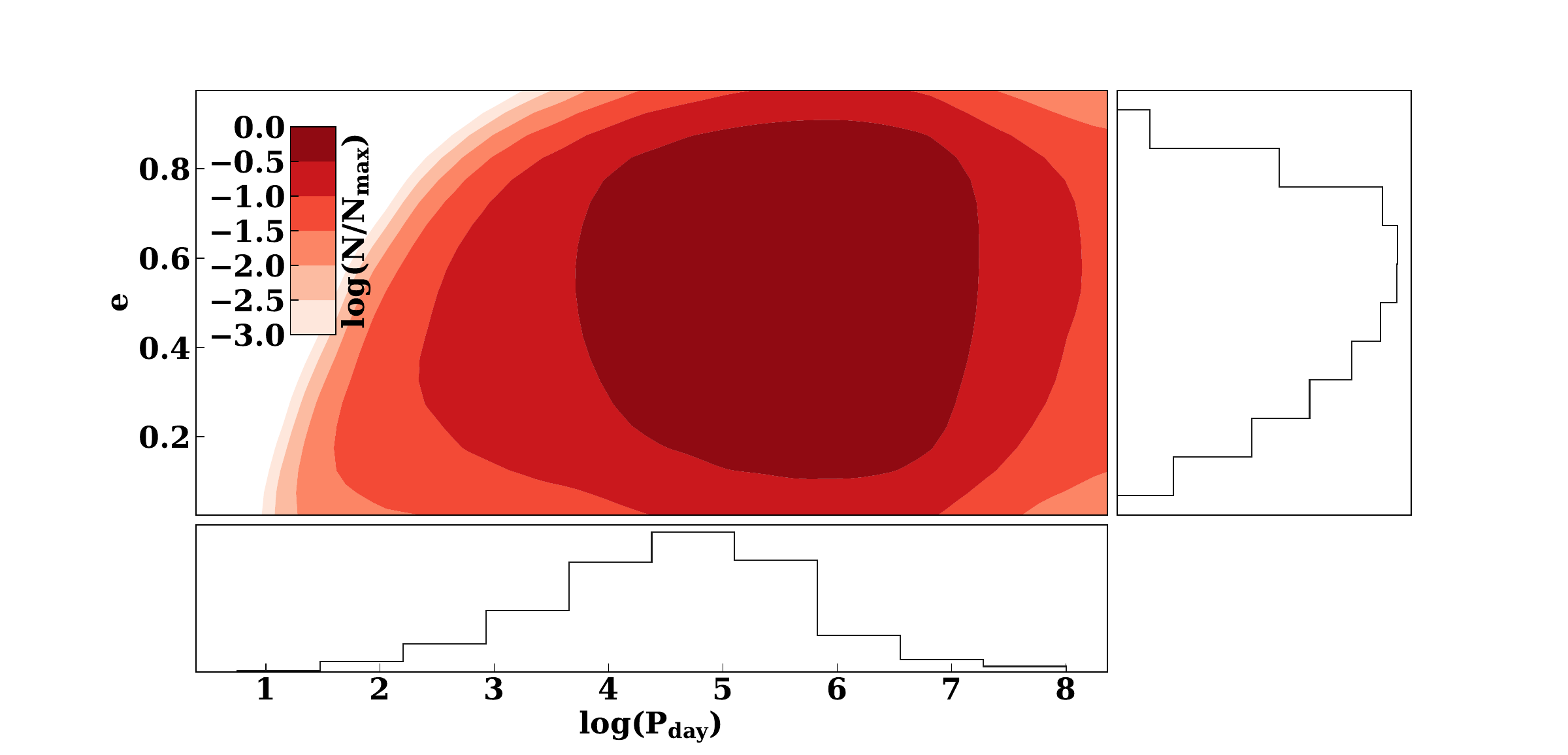}
    \figcaption{\label{fig:ica} Distribution of initial semi-major axes and eccentricities of triples' inner (top) and outer (bottom) orbit. The distributions are based on \citep{moe&distefano2017} (see text for details). The inner binary semi-major axis distribution is steeper than log-uniform primarily due to the \citet{mardling&aarseth2001} stability criterion.}
\end{figure*}

The distribution of initial orbital properties for the generated triples is shown in Figure~\ref{fig:ica}.

\subsection{Triple evolution}
We evolve the generated triples with MSE \citep{hamers+2021}. MSE models the evolution of hierarchical multiple systems (binaries, triple, quadruples, or higher order multiples), accounting for gravitational and binary interactions between the stars as well as stellar evolution. The code automatically switches between solving the secular equations of motion and direct N-body integration depending on the configuration of the system. Thus MSE can model both stable and unstable phases of evolution, and associated phenomenology (e.g. collisions, escapes, exchanges; see \citealp{toonen+2020,toonen+2022} for overviews of triple evolution).

We evolve each system for a random time between 10 Myr and 10 Gyr (implicitly assuming a constant star formation rate). We have updated the stellar wind, remnant mass, and supernova kick prescription in MSE, which followed the BSE
\citep{hurley+2002}, and were somewhat outdated, to match those in \citet{elbadry+2023}. After these updates, single-star evolution is consistent (at the few percent level) between the updated MSE and the binary population synthesis code COSMIC \citep{breivik+2020} used there.\footnote{The updated code is available at \url{https://github.com/alekseygenerozov/mse}. Please see the Update\_winds branch.}

As a control, we also evolve binaries generated via the procedure in \S~\ref{sec:methods} in isolation. Where possible, we match the binary and stellar evolution parameters of previous studies of Gaia BH1 \citep{elbadry+2023}, that used COSMIC. 

There are some differences in the treatment of binary evolution between MSE and COSMIC that are not easily removed. For example, COSMIC has different prescriptions for the onset of unstable mass transfer. On the other hand, MSE has prescriptions for eccentric mass transfer that are not in COSMIC (in particular the \citealt{hamers&dosopoulou2019} prescription that smoothly transitions from continuous mass transfer for circular orbits to impulsive mass transfer at pericenter for highly eccentric orbits). 

In this work, we set the common envelope efficiency ($\alpha$) to either 1 or 5 for simplicity. We also use the fall-back modulated kick prescription from \citet{fryer+2012}. In contrast, the common envelope efficiency and kick magnitude are free parameters in \citet{elbadry+2023}. 
For each set of parameters, we initialize $\sim 5\times 10^5$ systems.

MSE does not properly handle supernovae during 
common envelope phases, and we filter out such systems. This only occurs in a narrow region of parameters space with initially $\sim 20 M_{\odot}$ primaries that experience a common envelope during an AGB phase. Such primaries form $\sim 3 M_{\odot}$ BHs. This filter does not affect our conclusions on the formation of Gaia BH-like systems or on the overall formation BH binaries.

\section{Results}
\label{sec:results}
\begin{figure*}
    \plottwo{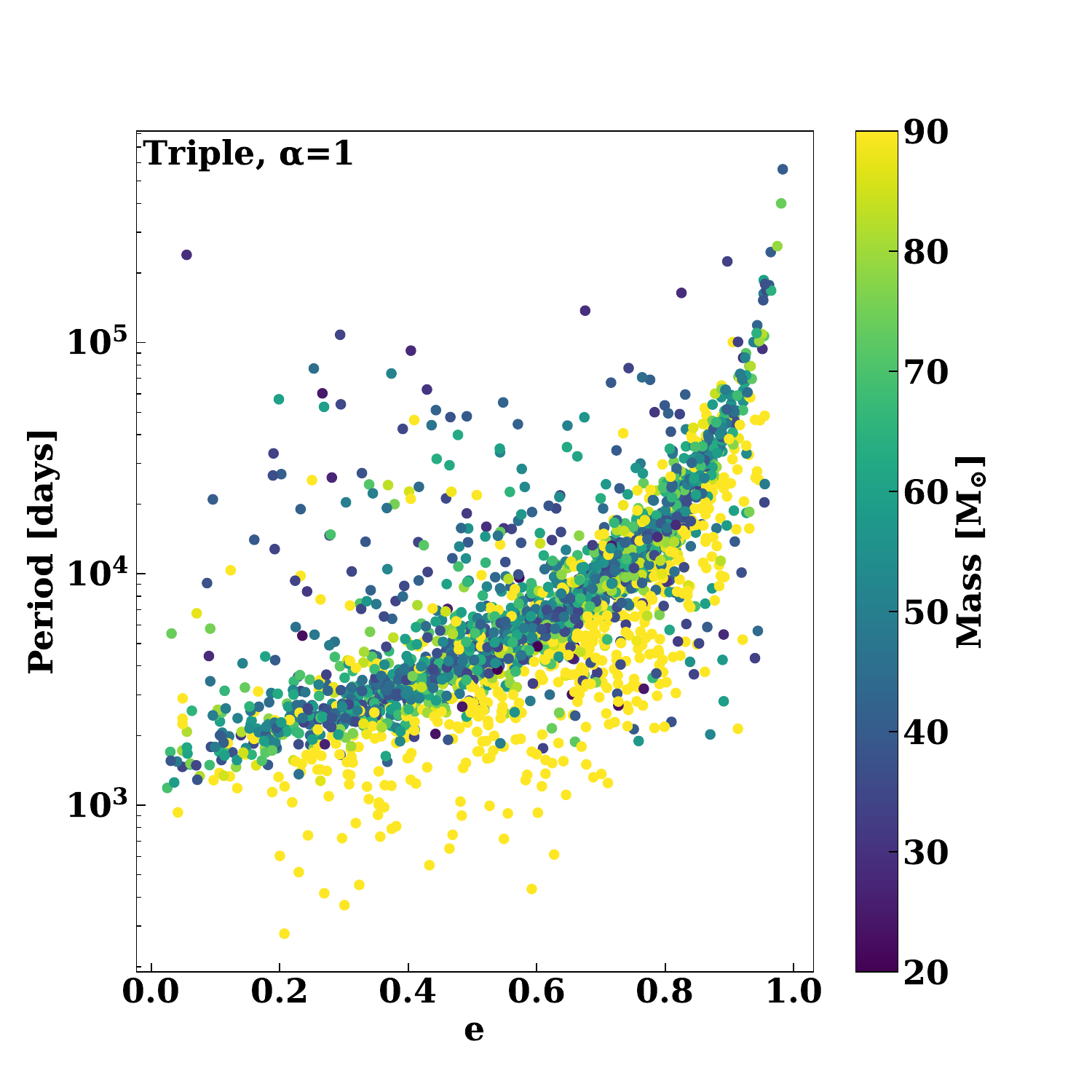}{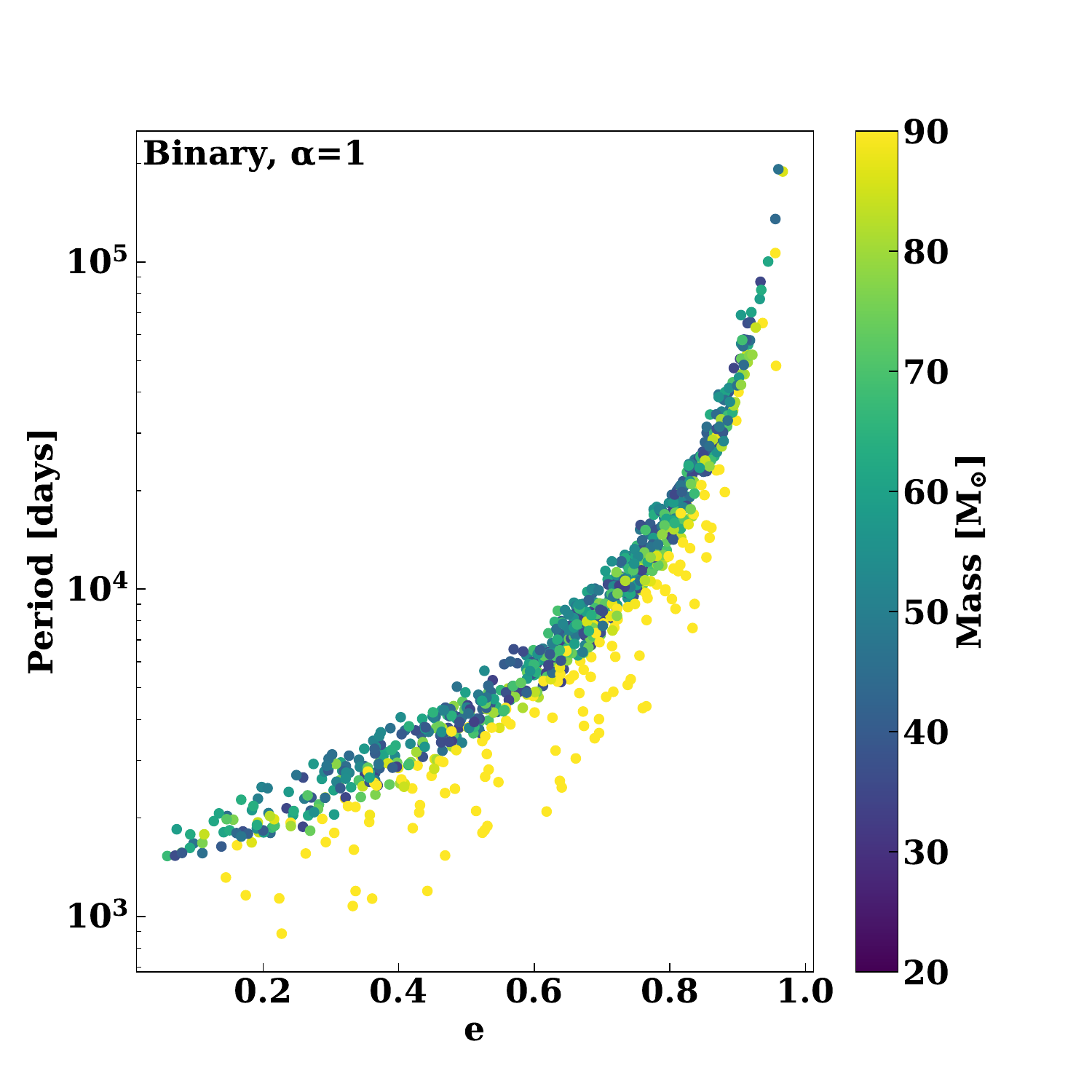}
    \plottwo{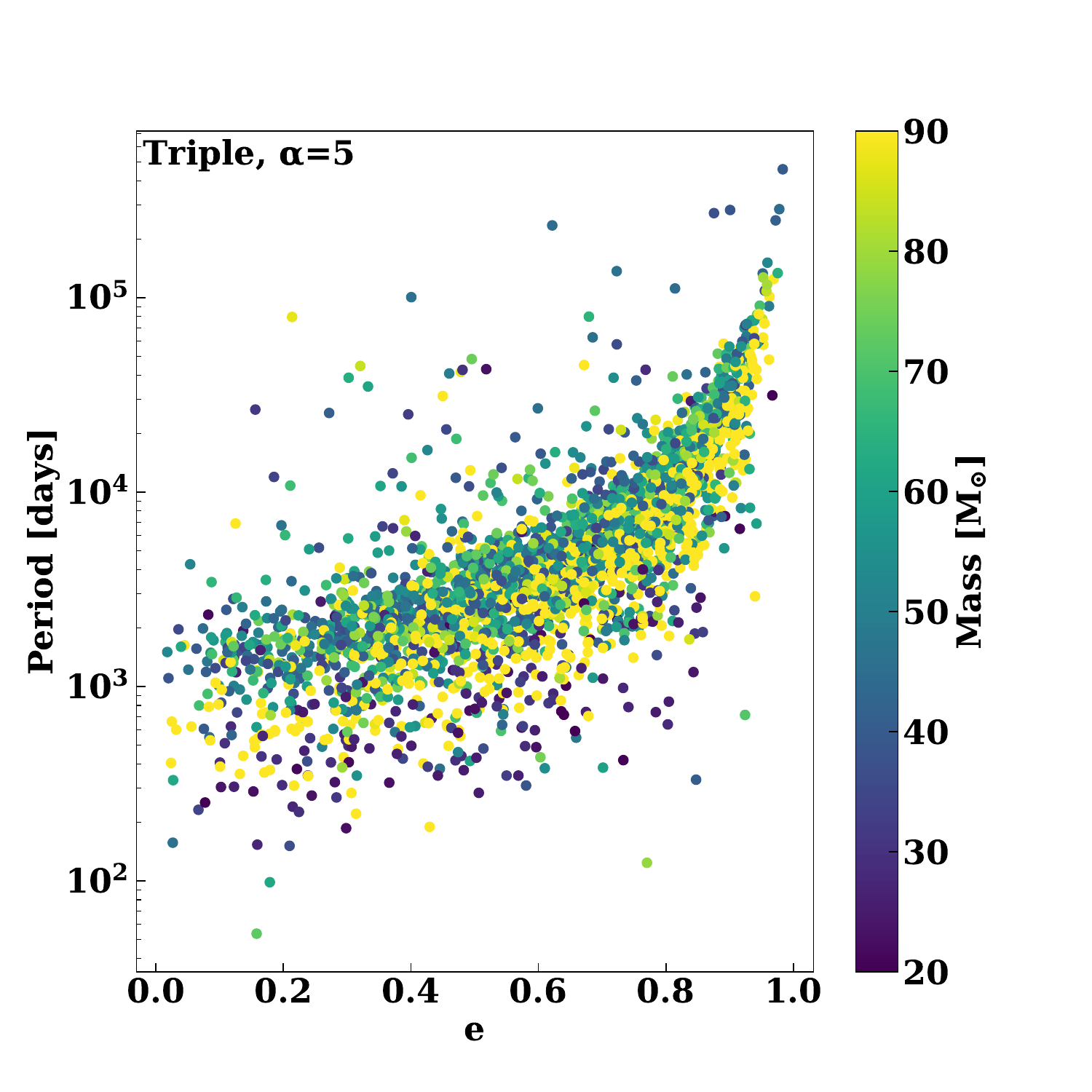}{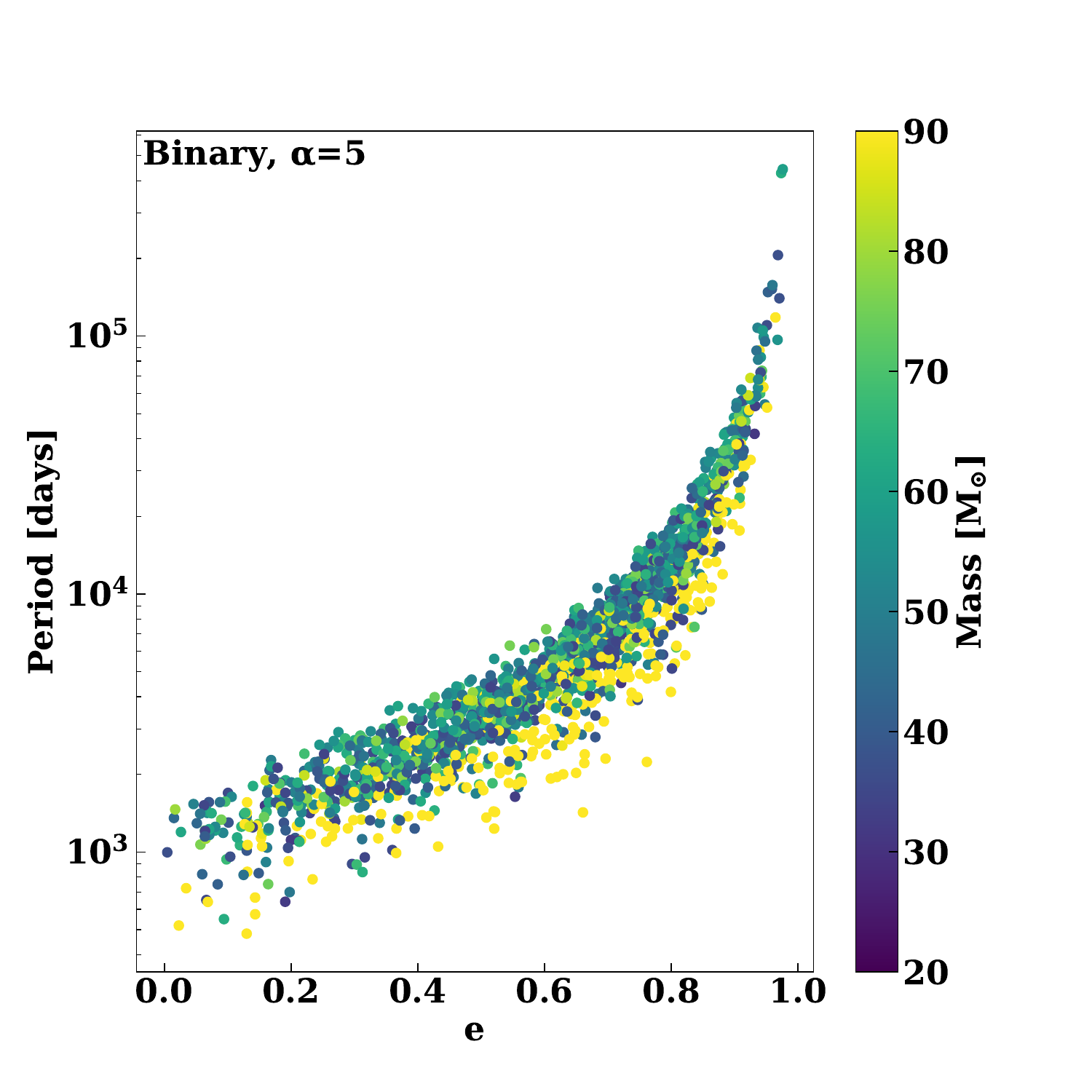}
    \figcaption{\label{fig:tripic1} Initial periods and eccentricities for stars that end up in BH-main sequence binaries with periods $<$ 1300 days after triple (top) and binary evolution (bottom). For triples, the 
    stellar companion was initially the tertiary star 80\% of the time.}
\end{figure*}

\begin{figure*}
    \plottwo{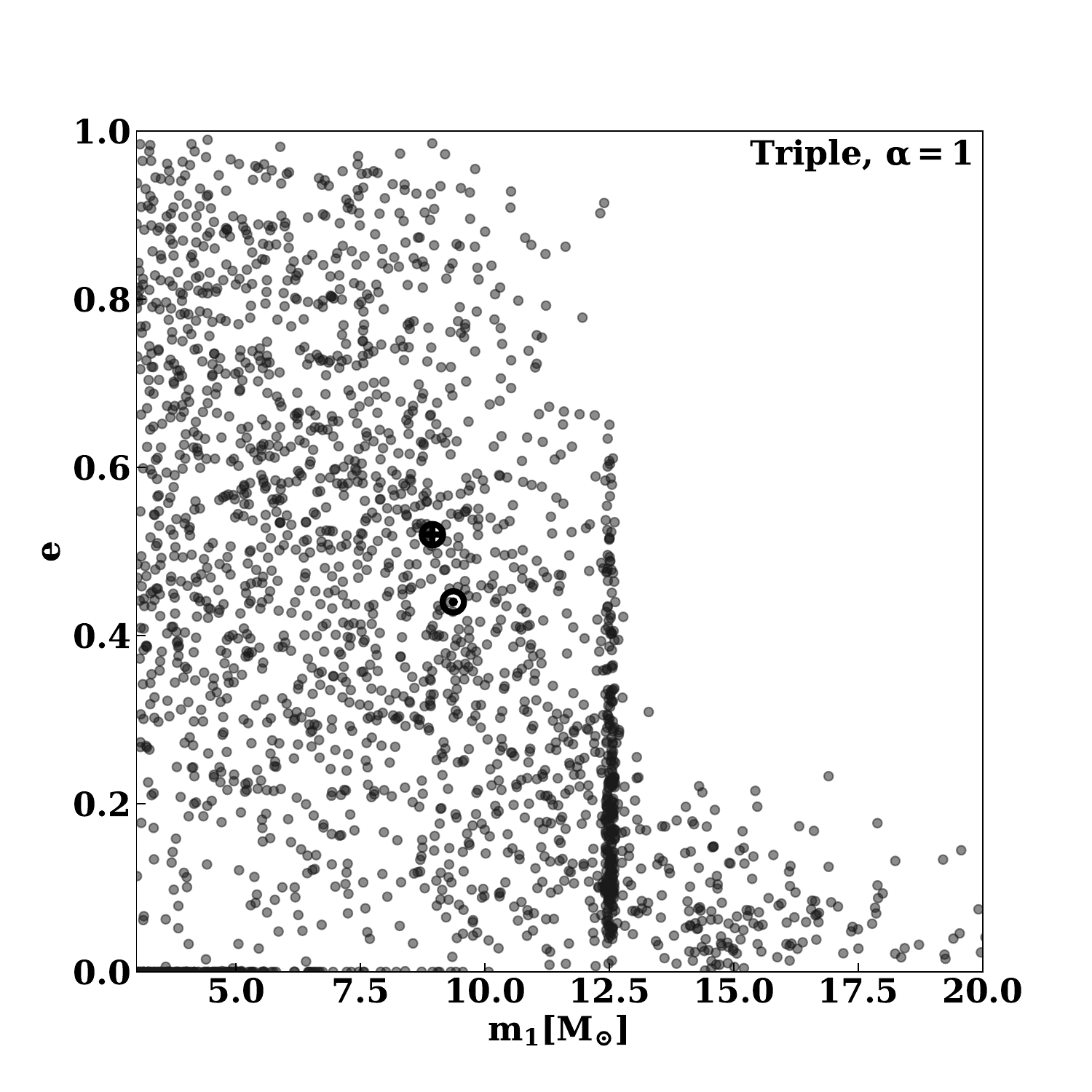}{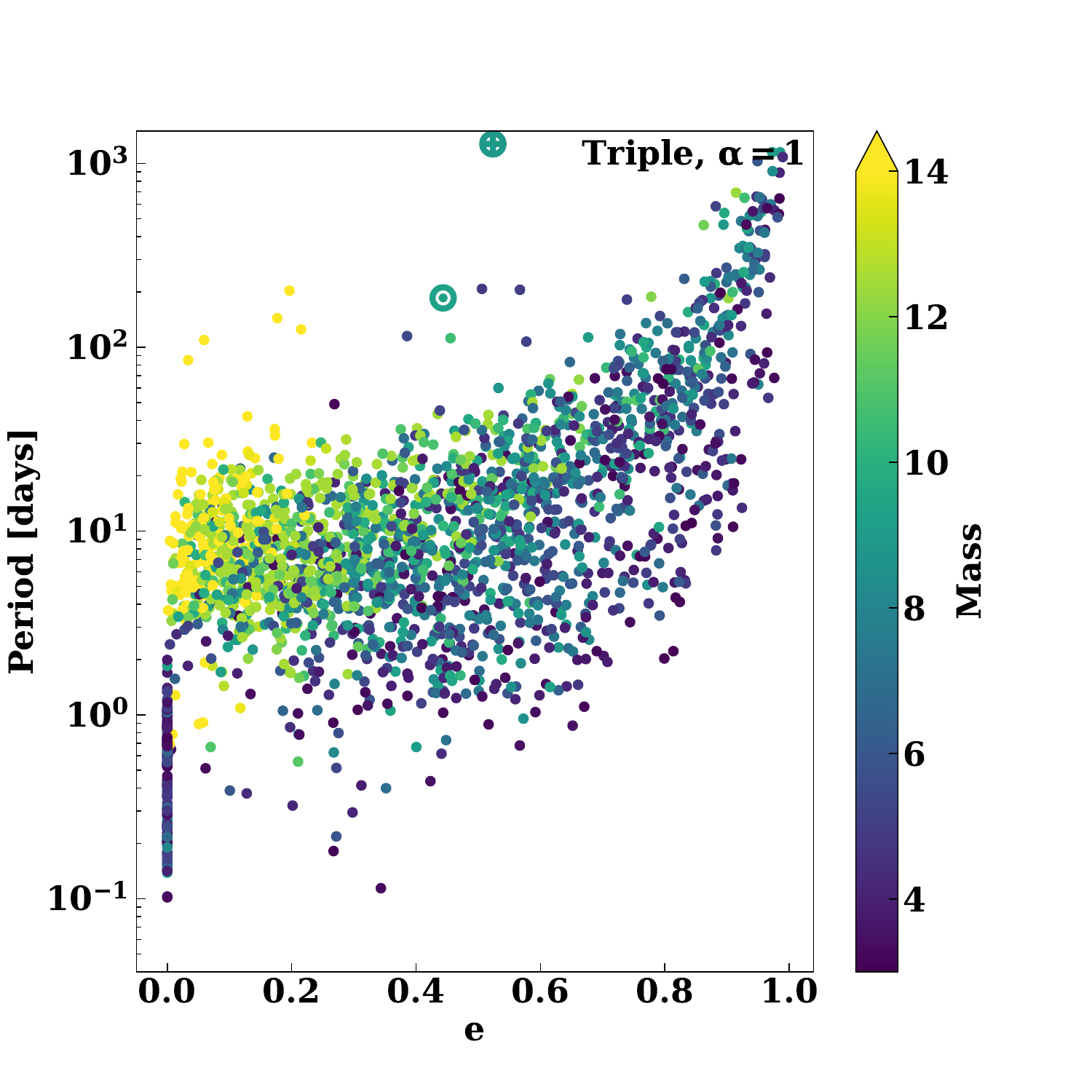}
    \plottwo{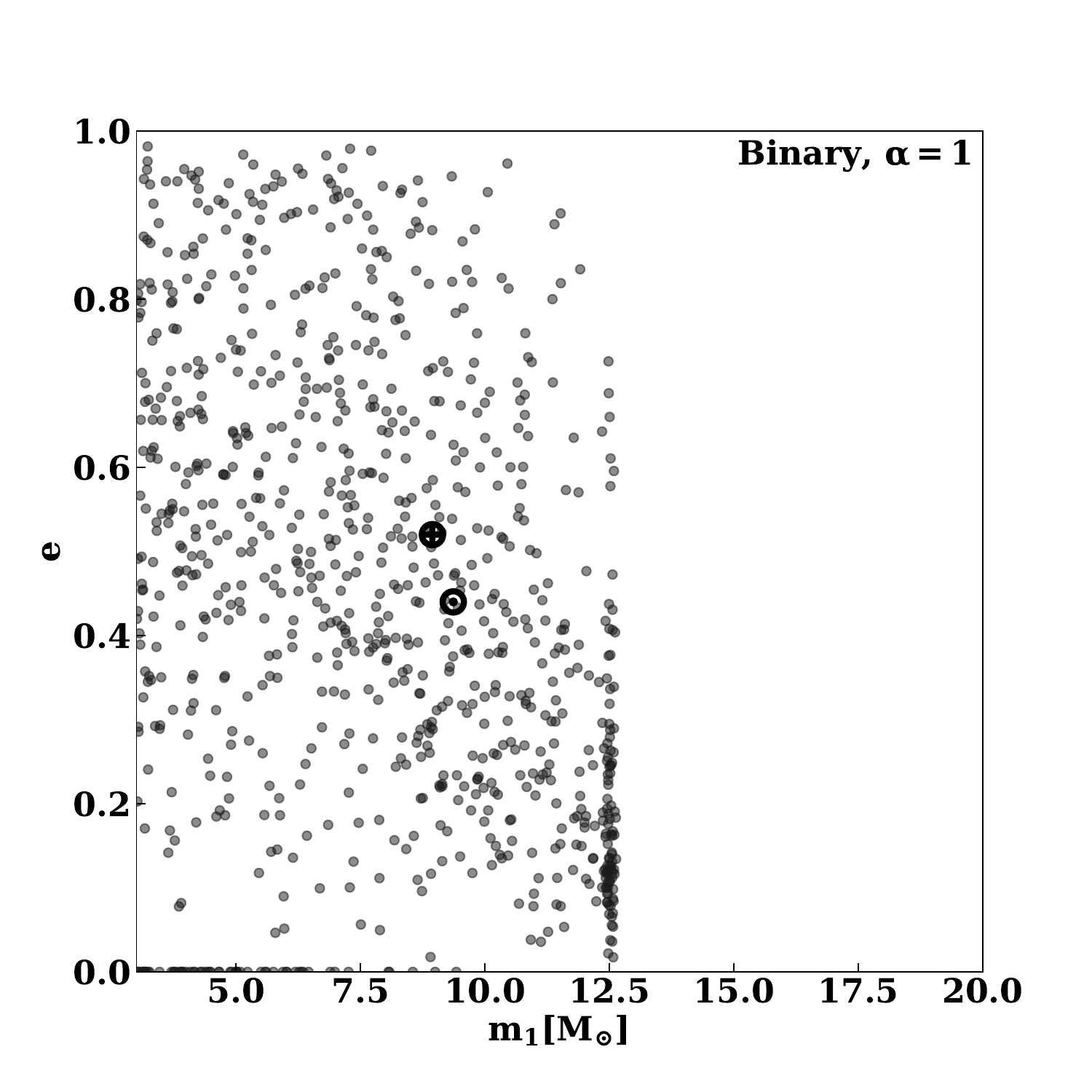}{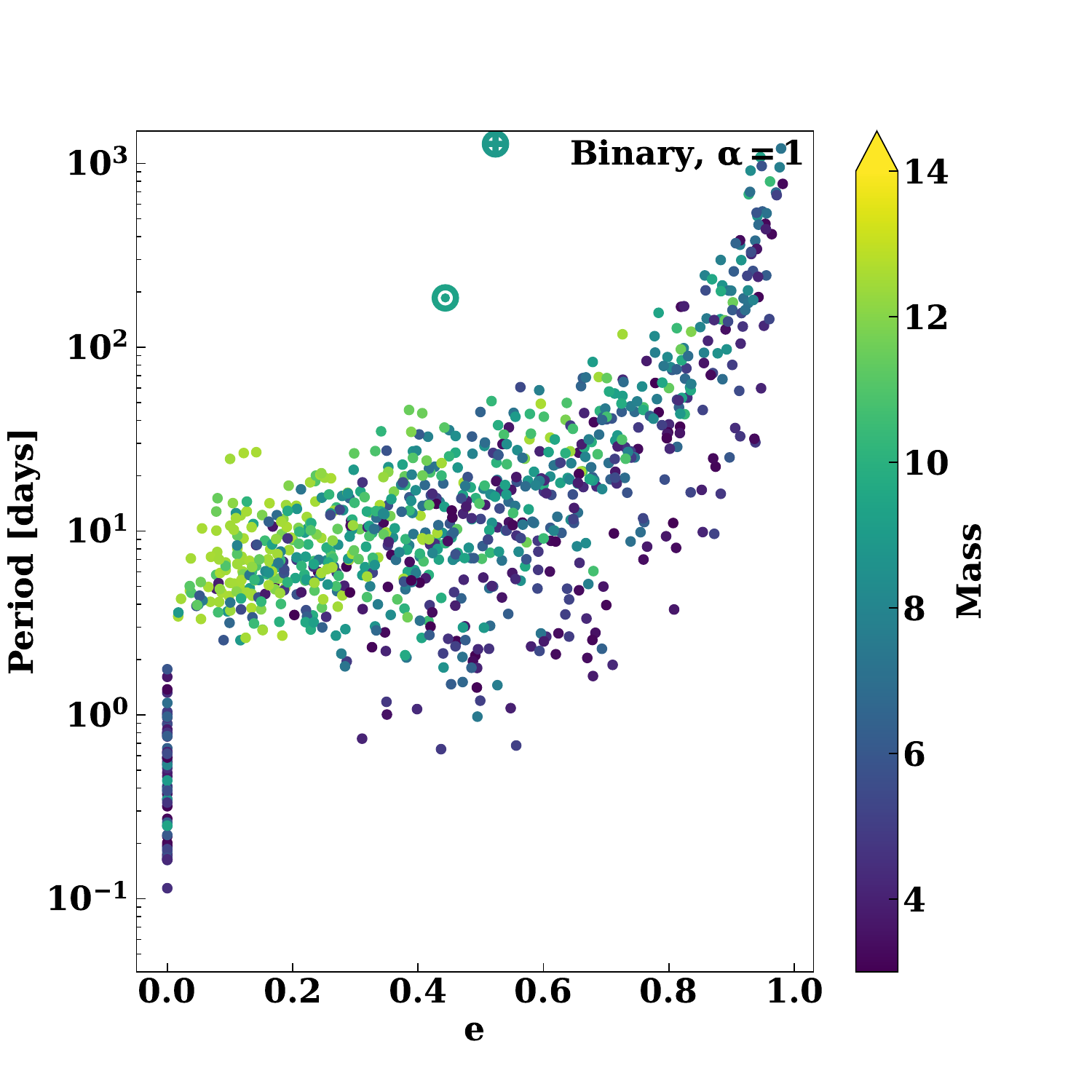}
    \figcaption{\label{fig:tripRes1} Left panels: Eccentricity as a function of primary mass for surviving BH-main sequence binaries with periods $<$ 1300 days after triple (top) and binary evolution (bottom). Right panels: Period as a function of eccentricity for surviving BH-main sequence binaries. The circle dot (plus) symbol is Gaia BH1 (BH2). Unlike the simulated systems here, Gaia BH2 has a red giant companion}
\end{figure*}

\begin{figure*}
    \plottwo{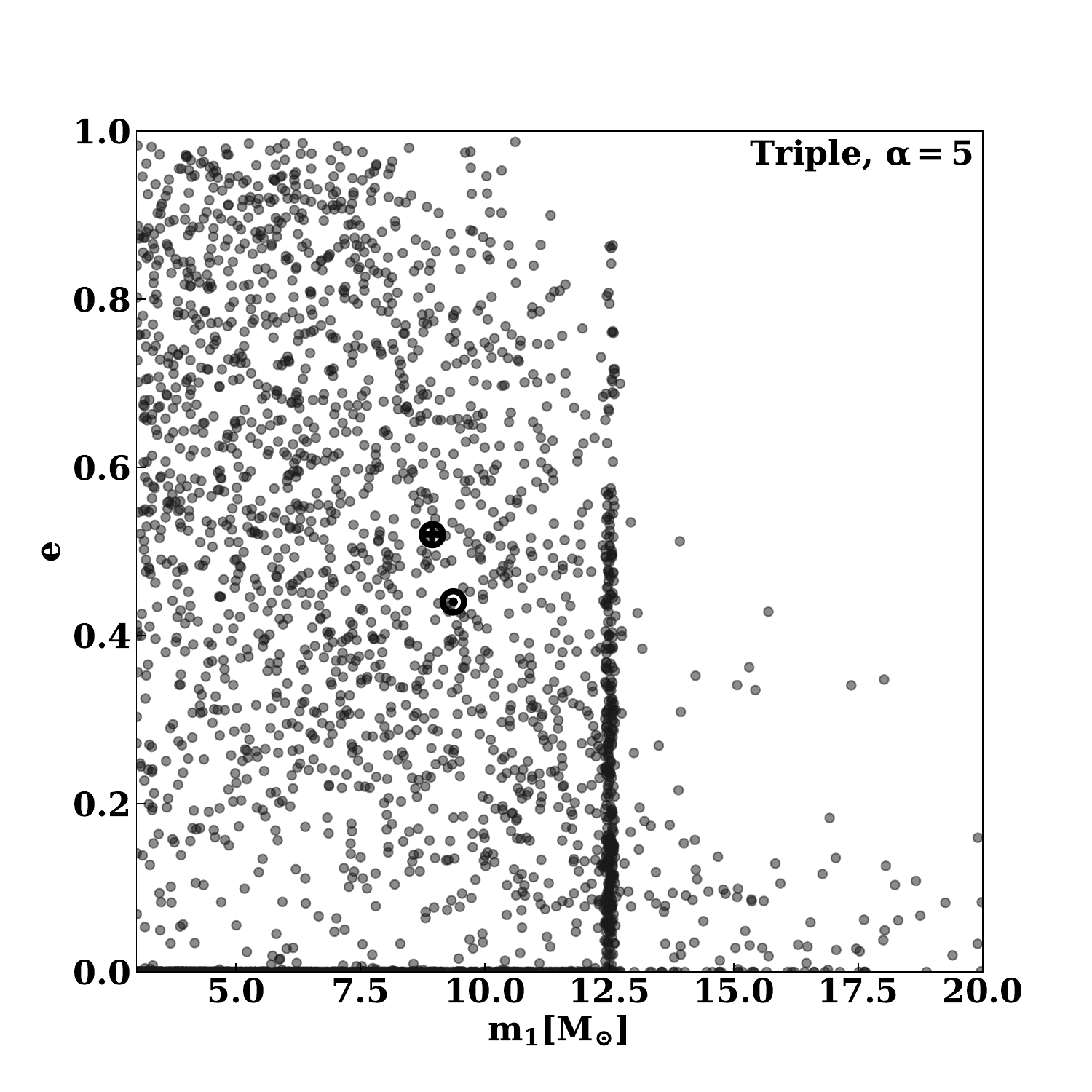}{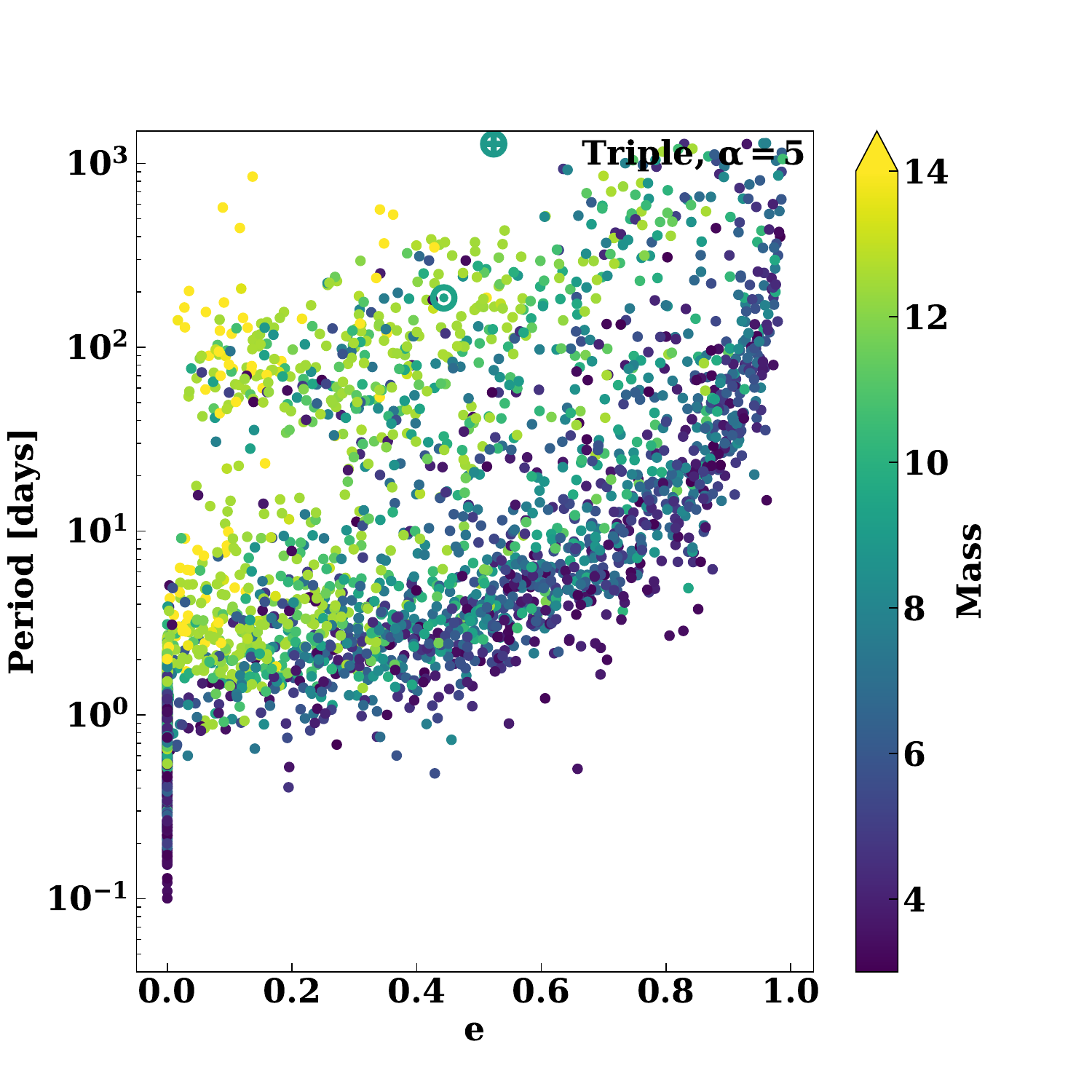}
    \plottwo{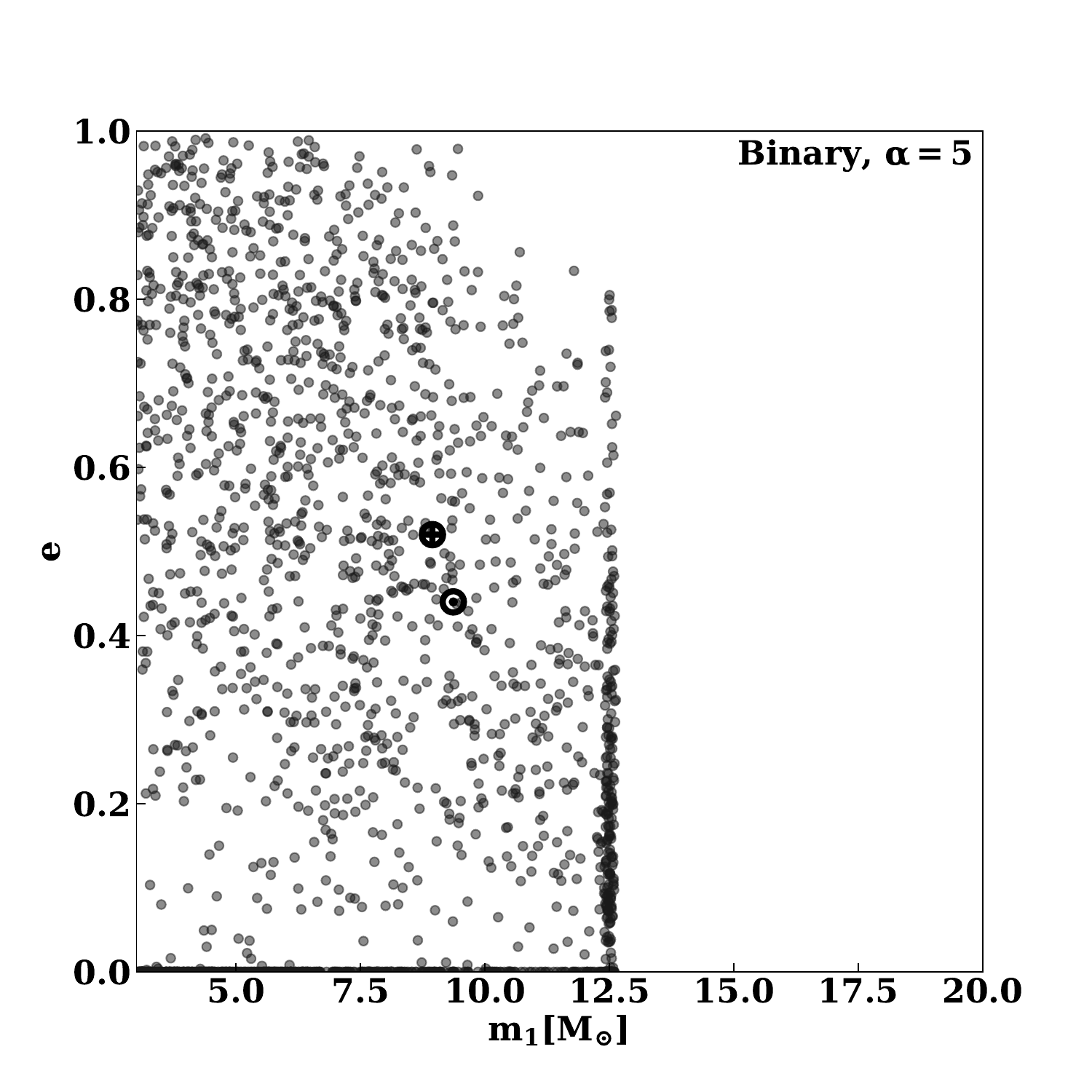}{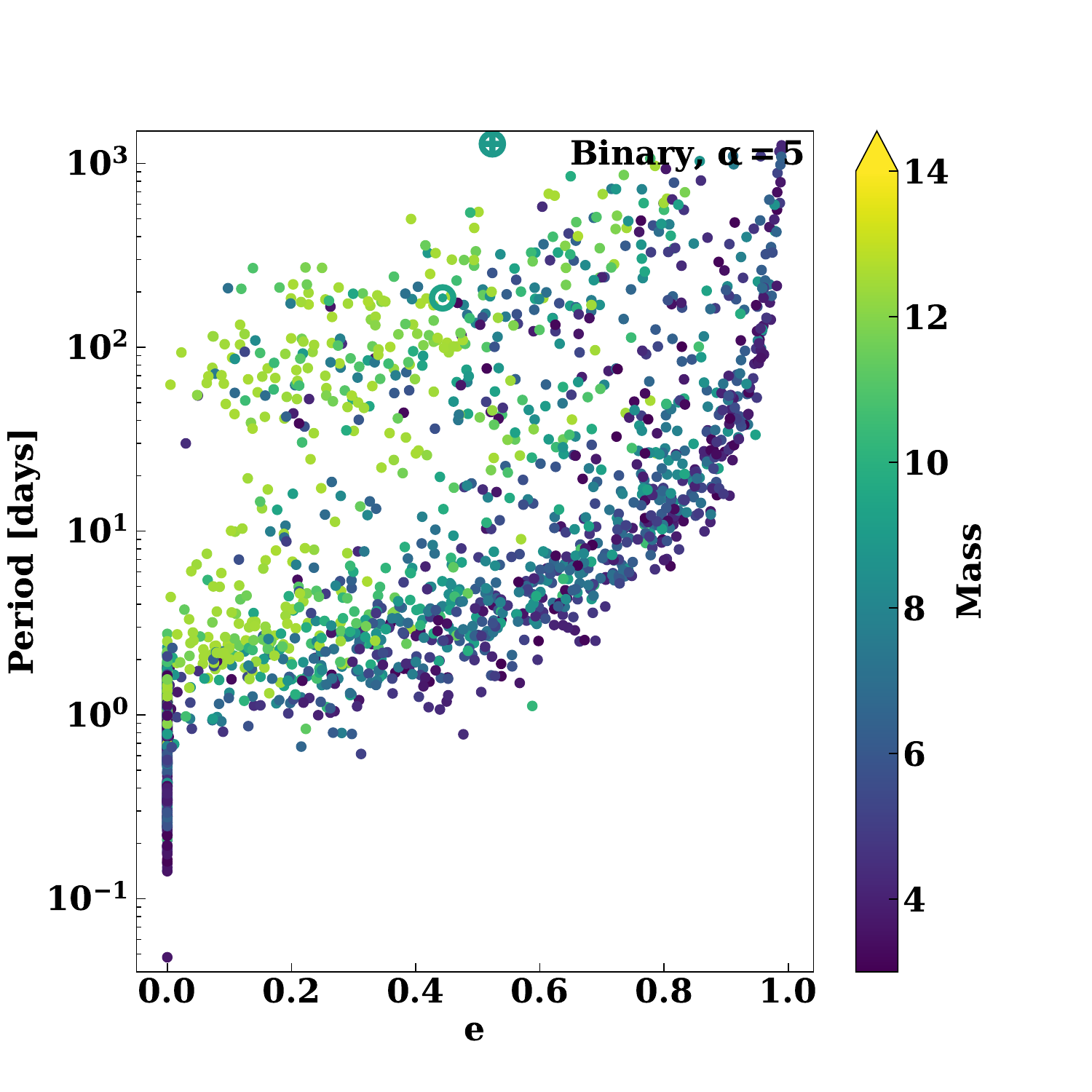}
    \figcaption{\label{fig:tripRes2} Same as Figure~\ref{fig:tripRes1}, except for a common envelope efficiency ($\alpha$) of 5.}
\end{figure*}

The top panels of Figure~\ref{fig:tripic1} show the initial period and eccentricity of stars that end up in BH-main sequence binaries with periods $< 1300$ days, following triple evolution.
The top, left panels of Figure~\ref{fig:tripRes1} and Figure~\ref{fig:tripRes2} show the final BH masses and eccentricities for these systems. The top, right panels show the final period as a function of eccentricity. The binaries in Figures~\ref{fig:tripRes1} and~\ref{fig:tripRes2} are no longer in triples, but the stellar companion in these systems was the tertiary star of the original triple $\sim 80\%$ of the time. 
The bottom panels of Figures~\ref{fig:tripic1} and Figures~\ref{fig:tripRes1} and~\ref{fig:tripRes2} show the same distributions for binaries evolved in isolation. 

For low common envelope efficiencies ($\alpha=1$), binary evolution is unable to produce systems with periods $\gsim 100$ days and moderate eccentricities ($e \lsim 0.7$) like Gaia BH1. Triple evolution, on the other hand, can produce such systems. 

\begin{figure*}
    \includegraphics[width=\textwidth]{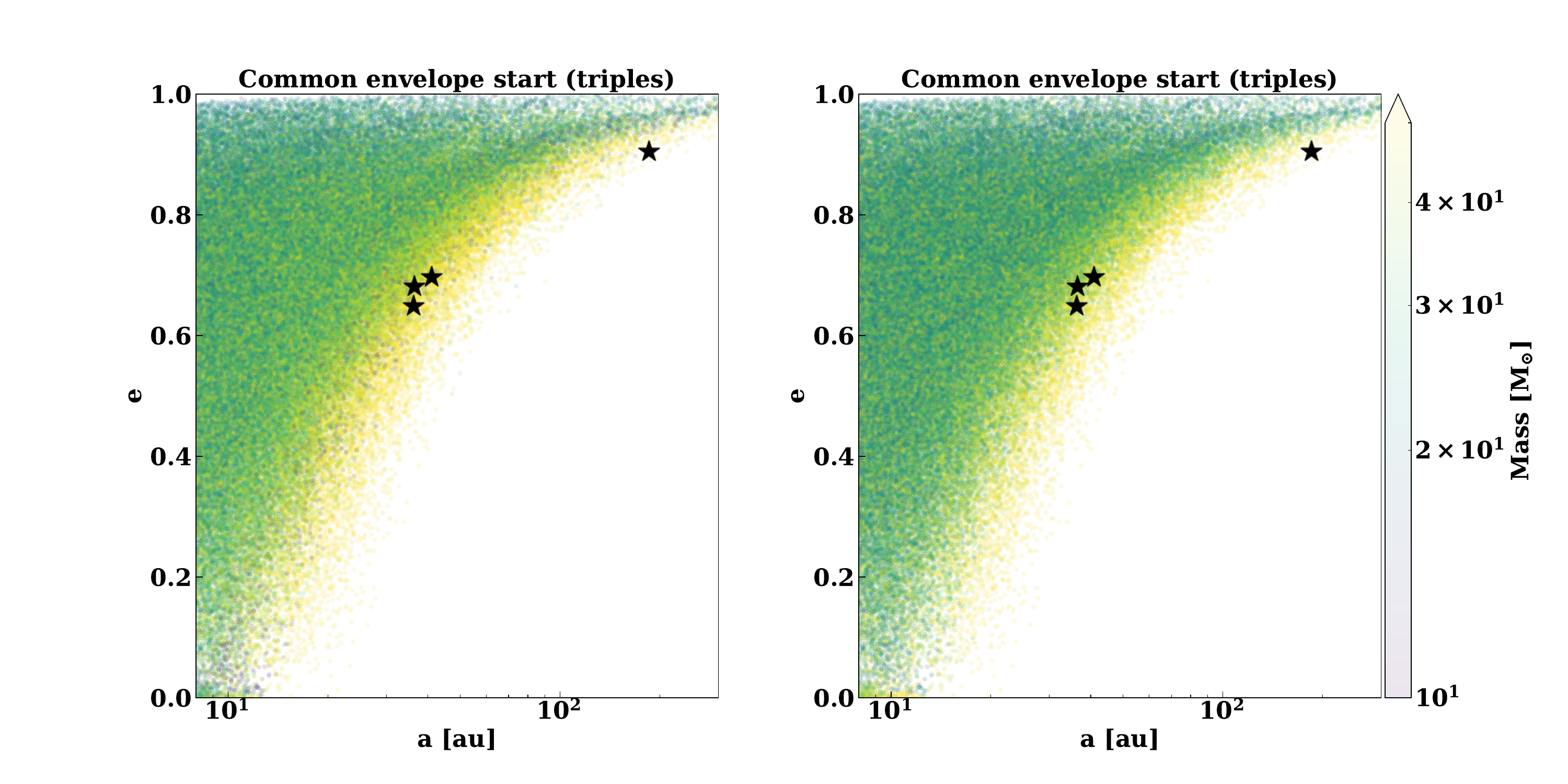}
    \includegraphics[width=\textwidth]{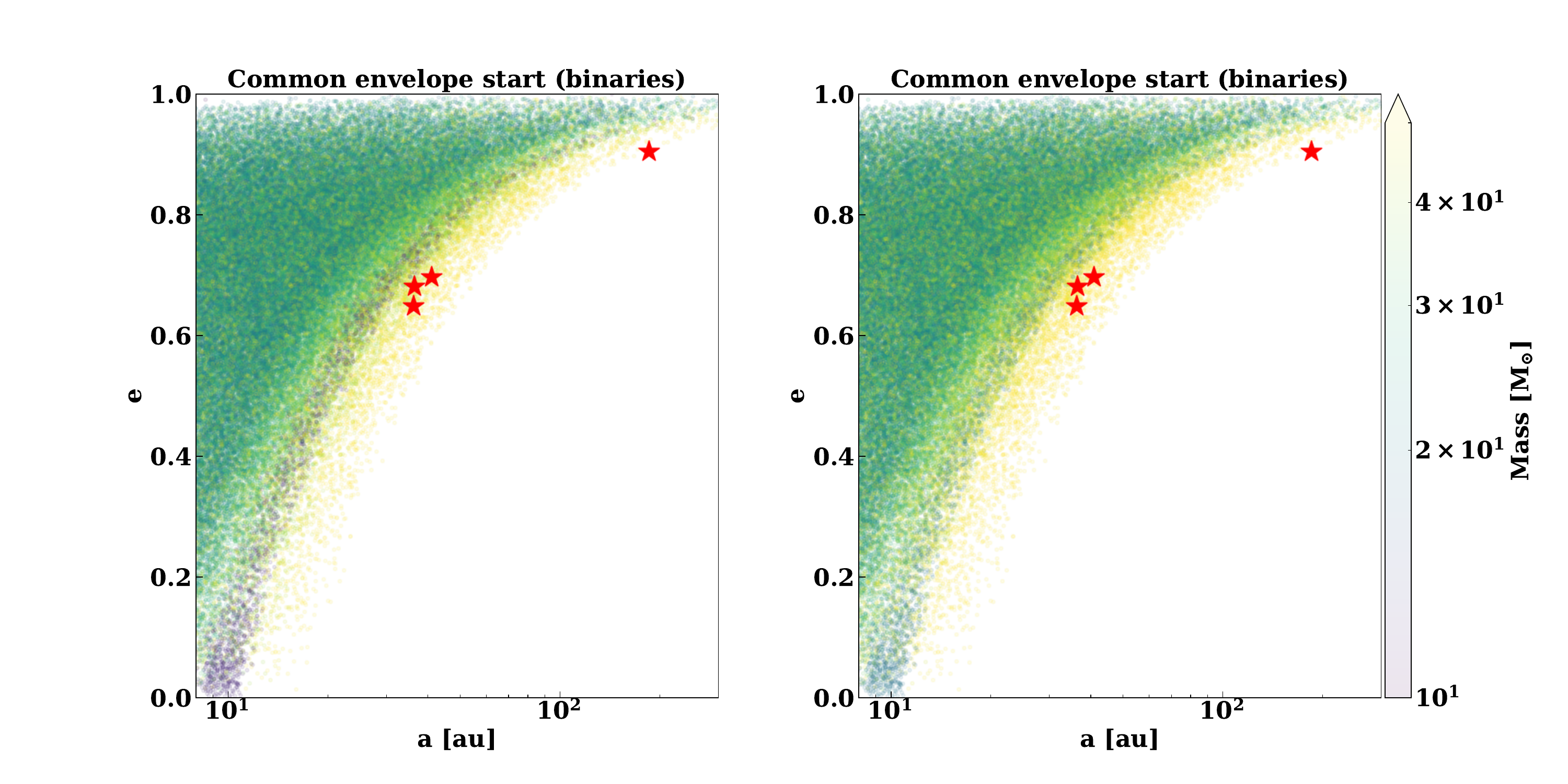}
    \figcaption{\label{fig:ce} Eccentricity as a function of semi-major axis at the start of the common envelope phase for triples (first row) and binaries (second row). For triples, we show the orbital elements of the innermost binary at the last common envelope phase. The colors indicate the primary mass at the beginning of the common envelope (left) and at the initial condition (right). The black stars correspond to triples that end their evolution as Gaia BH1-like binaries (see text for definition). For comparison, we also reproduce the starred points from the top panel in red in the bottom panel.
    }
\end{figure*}

\begin{figure}
    \plotone{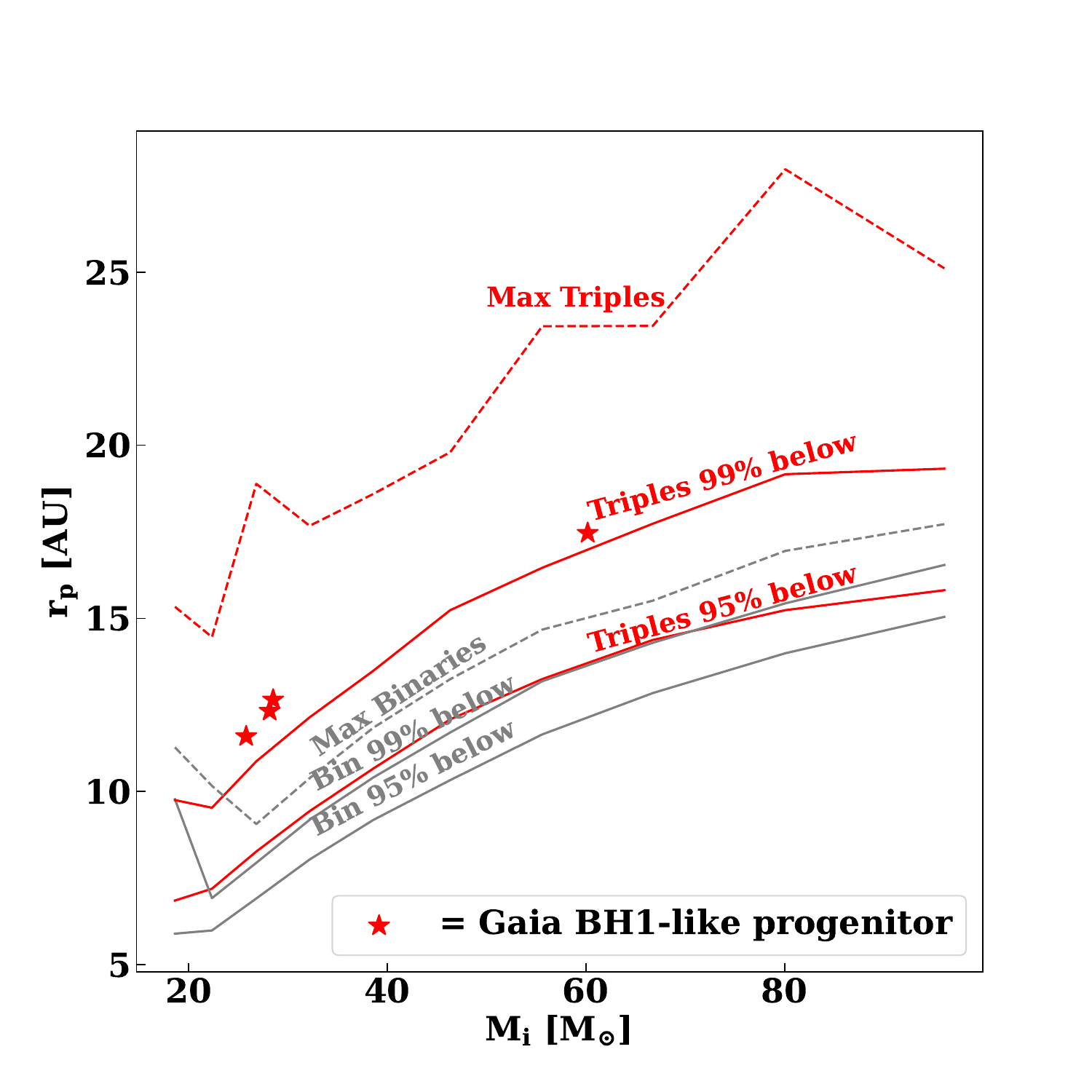}
    \figcaption{\label{fig:ce2} Maximum pericenter at the start of the common envelope as a function of the initial primary mass for binaries (dashed, gray) and triples (red, dashed). The red stars correspond to the model triple systems that form Gaia BH1-like binaries. The solid contours enclose 95\%-99\% of model triples or binaries.
    }
\end{figure}

\begin{figure*}
    \epsscale{0.8}
    \plotone{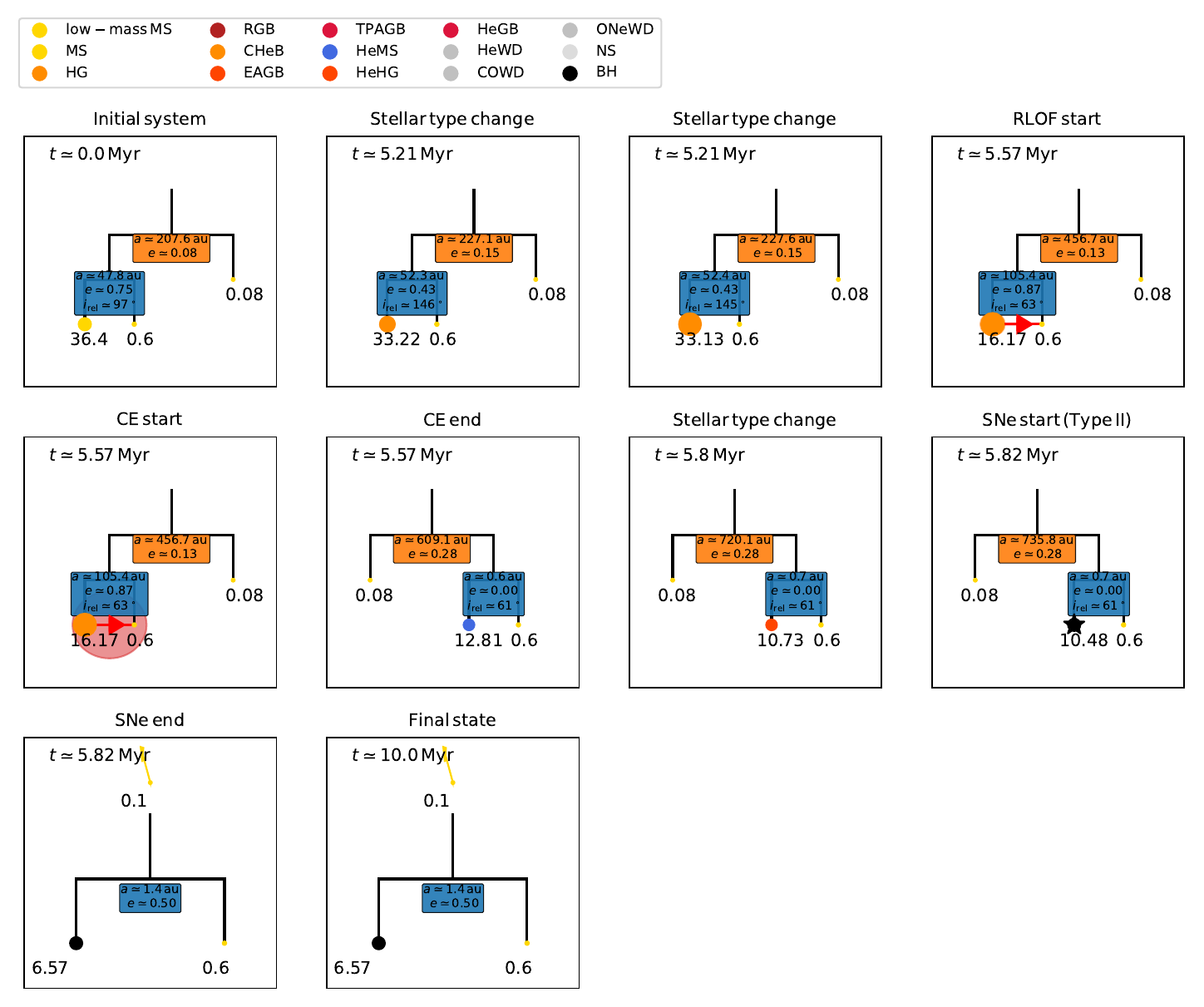}
    \plotone{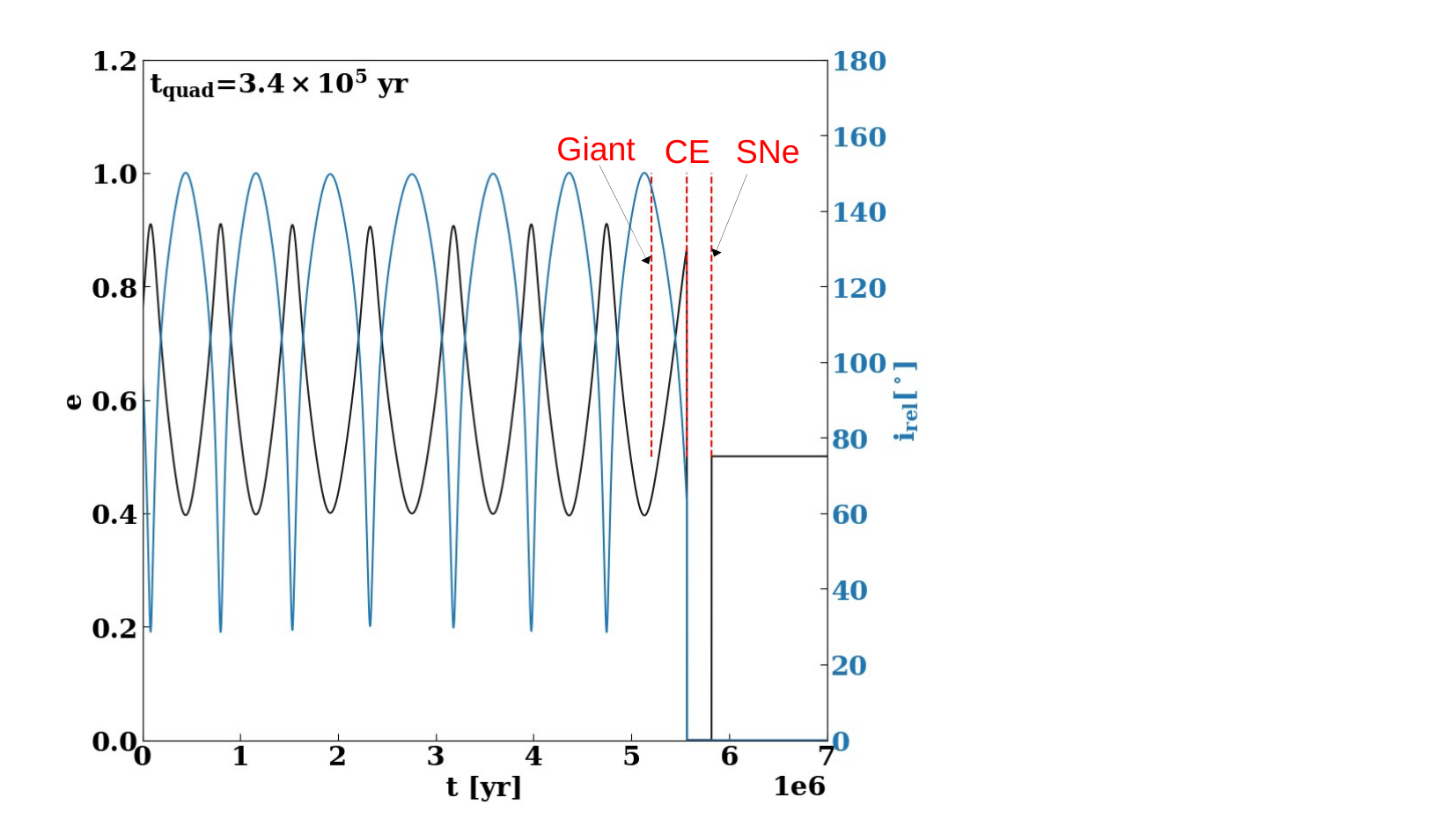}
    \caption{\label{fig:evolve_example} Example triple evolution that results in a Gaia BH1-like binary. The top panel shows the evolution of stars and binaries with 
    mobile diagrams. The bottom panel shows the evolution of the eccentricity and inclination as a function of time. These elements evolve due to 
    the von Zeipel--Lidov--Kozai effect. For reference, we show the quadrupole Kozai timescales in the bottom panel \citep{antognini2015,naoz+2016}.}
\end{figure*}

\begin{figure*}
    \epsscale{0.8}
    \plotone{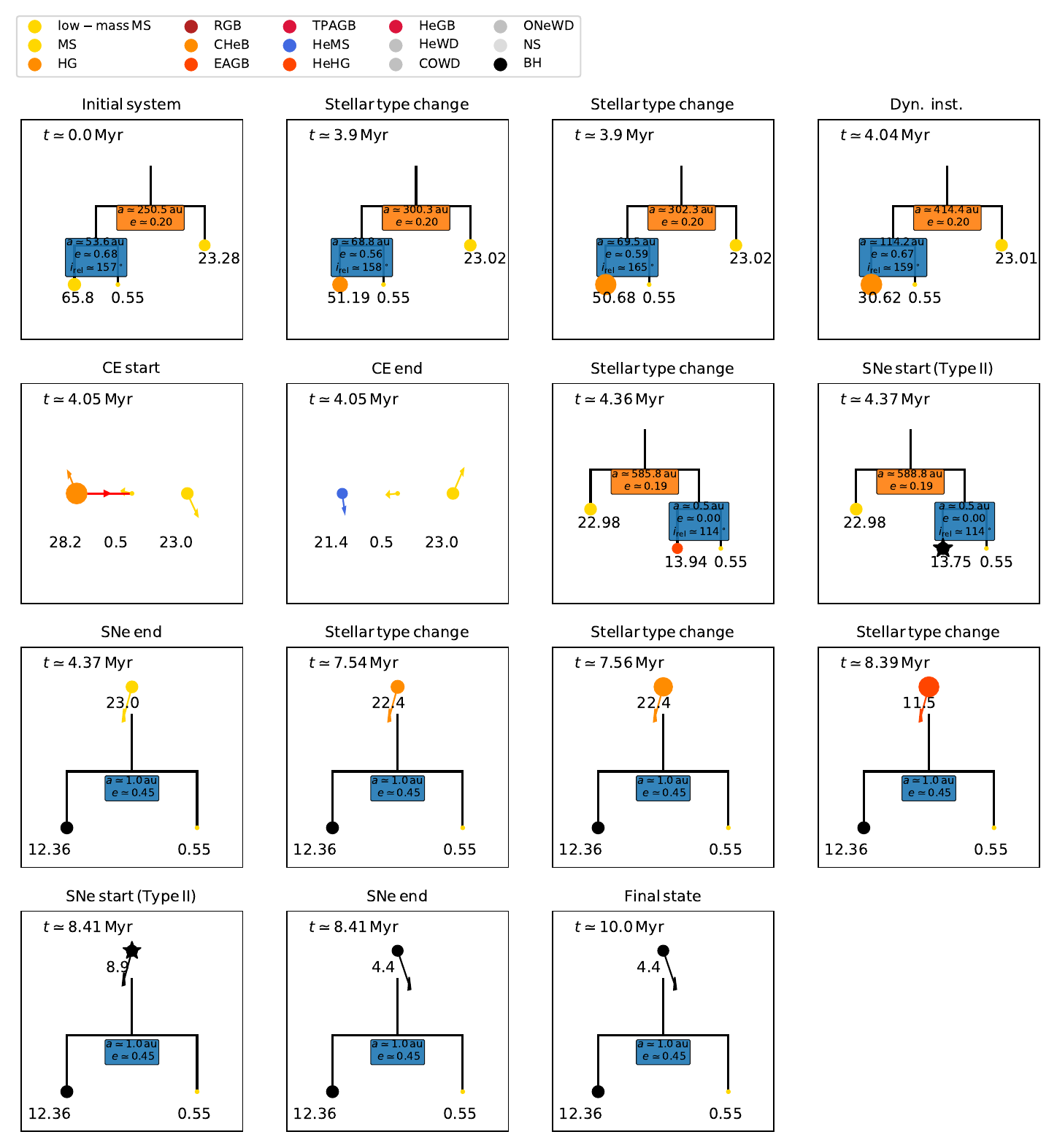}
    \plotone{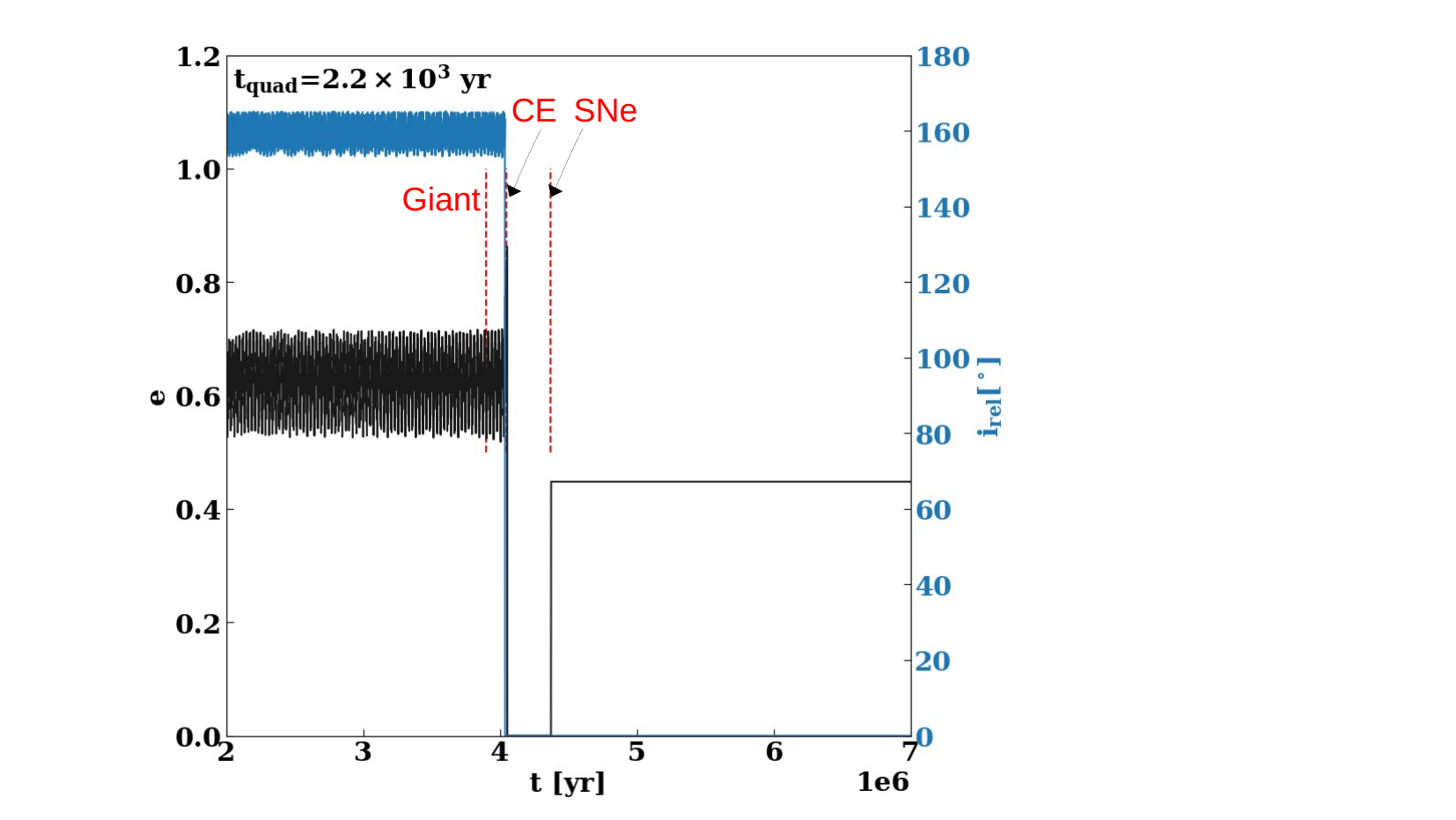}
    \figcaption{\label{fig:evolve_exampleTEDI} Same as Figure~\ref{fig:evolve_example}, except in this case, Gaia BH1-like binary is formed after a triple evolution dynamical instability (TEDI).}
\end{figure*}

\begin{figure*}
    \epsscale{0.8}
    \plotone{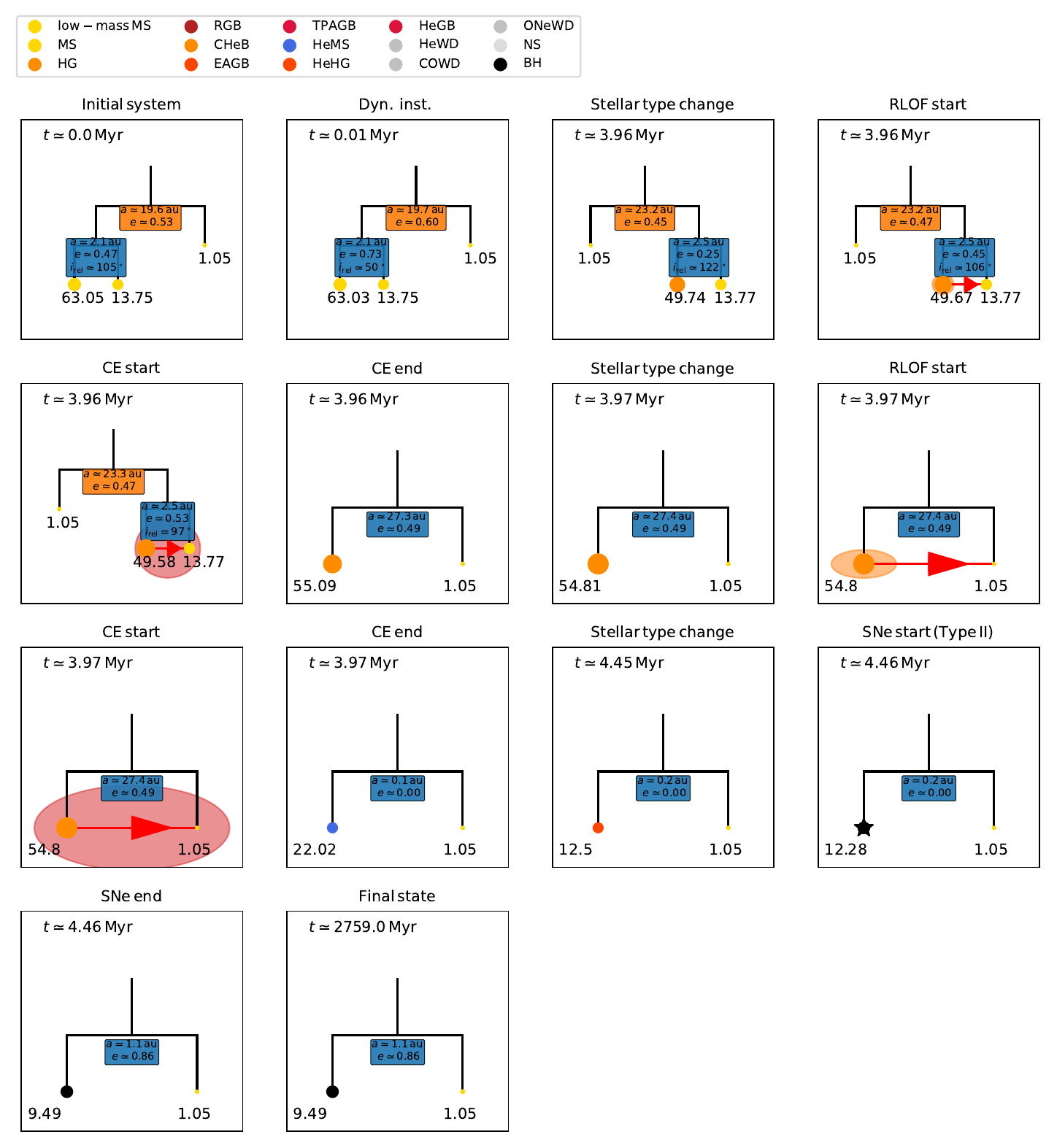}
    \figcaption{\label{fig:evolve_example2} Formation of a BH-main sequence binary, following two common-envelope phases.}
\end{figure*}

Gaia BH1-like systems form from the initial primary and secondary star, after a common-envelope phase. (We define a Gaia BH1-like system to be one where the eccentricity is between 2/3 and 3/2 times the observed one, while the BH mass, secondary mass, and period are between half and twice the observed ones).  In binaries, the separation before the common envelope only evolves adiabatically due to wind mass loss (with constant eccentricity). Therefore, the common envelope occurs at the beginning of the giant phase. Afterward, the separation increases much faster than the radius due to mass loss from the giant star.  In triples, the eccentricity of the secondary can evolve via the von Zeipel--Kozai--Lidov effect \citep{kozai1962,lidov1962,naoz+2016}.  Thus, in some cases, the eccentricity can decrease at the beginning of the giant phase, and then increase again so that the common envelope occurs later when the separation between the low-mass star and BH progenitor is wider, as illustrated in Figures~\ref{fig:ce},~\ref{fig:ce2}, and~\ref{fig:evolve_example}. Moreover, stellar evolution and mass loss can trigger and change such secular evolution \citep{perets&kratter2012,Mic+14}. Alternatively, the eccentricity of the secondary can be excited by a triple evolution dynamical instability (TEDI) \citep{perets&kratter2012, Mic+14, toonen+2020,hamers+2022}. In this case, the stellar eccentricity is excited, as the system becomes less hierarchical due to stellar evolution. This type of evolution is shown in Figure~\ref{fig:evolve_exampleTEDI}.\footnote{The examples in Figures~\ref{fig:evolve_example}--~\ref{fig:evolve_example2} are from a separate batch of simulations run for $10^7$ yr with high output cadence to ensure well-sampled orbital data.}

Other formation channels are possible for broader definitions of Gaia BH1-like binaries. For example, allowing the eccentricity and other properties to differ by a factor of two from Gaia BH1. This modified definition picks out a distinct binary population with eccentricity $\gsim 0.7$. In most cases ($68\%$ for $\alpha=1$; $81\%$ for $\alpha=5$), such binaries are formed after a stellar merger in the inner binary followed by a common envelope with the initial tertiary star. This evolution is illustrated in Figure~\ref{fig:evolve_example2}.  This behavior depends on uncertain prescriptions for the stellar radius of high-mass stars. In fact, high-mass stars ($\gsim 50 M_{\odot}$) may not enter a red giant phase (see \S~\ref{sec:alt}). Indeed, it was already suggested that stellar mergers in triples can produce blue straggler binaries \citep{perets&fabrycky2009}.

\citet{rastello+2023}, propose an alternative definition of Gaia BH1-like binaries with more stringent requirements on the component masses. In particular, they require the BH to be $8-12 M_{\odot}$ and the companion to be $0.5 - 1.5 M_{\odot}$. With this definition, only one Gaia BH1-like system is produced for $\alpha=1$.

\begin{figure*}
    \includegraphics[width=\columnwidth]{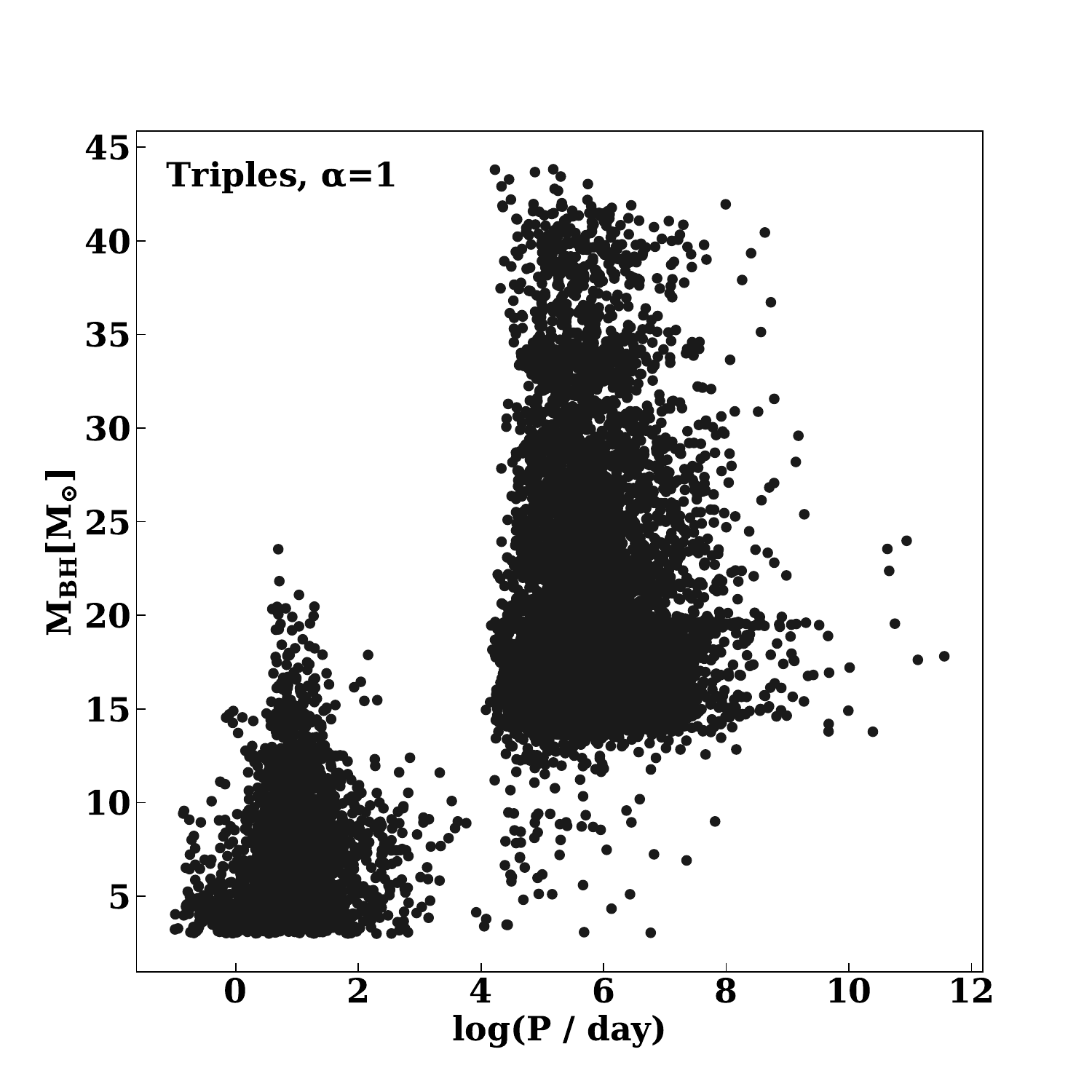}
    \includegraphics[width=\columnwidth]{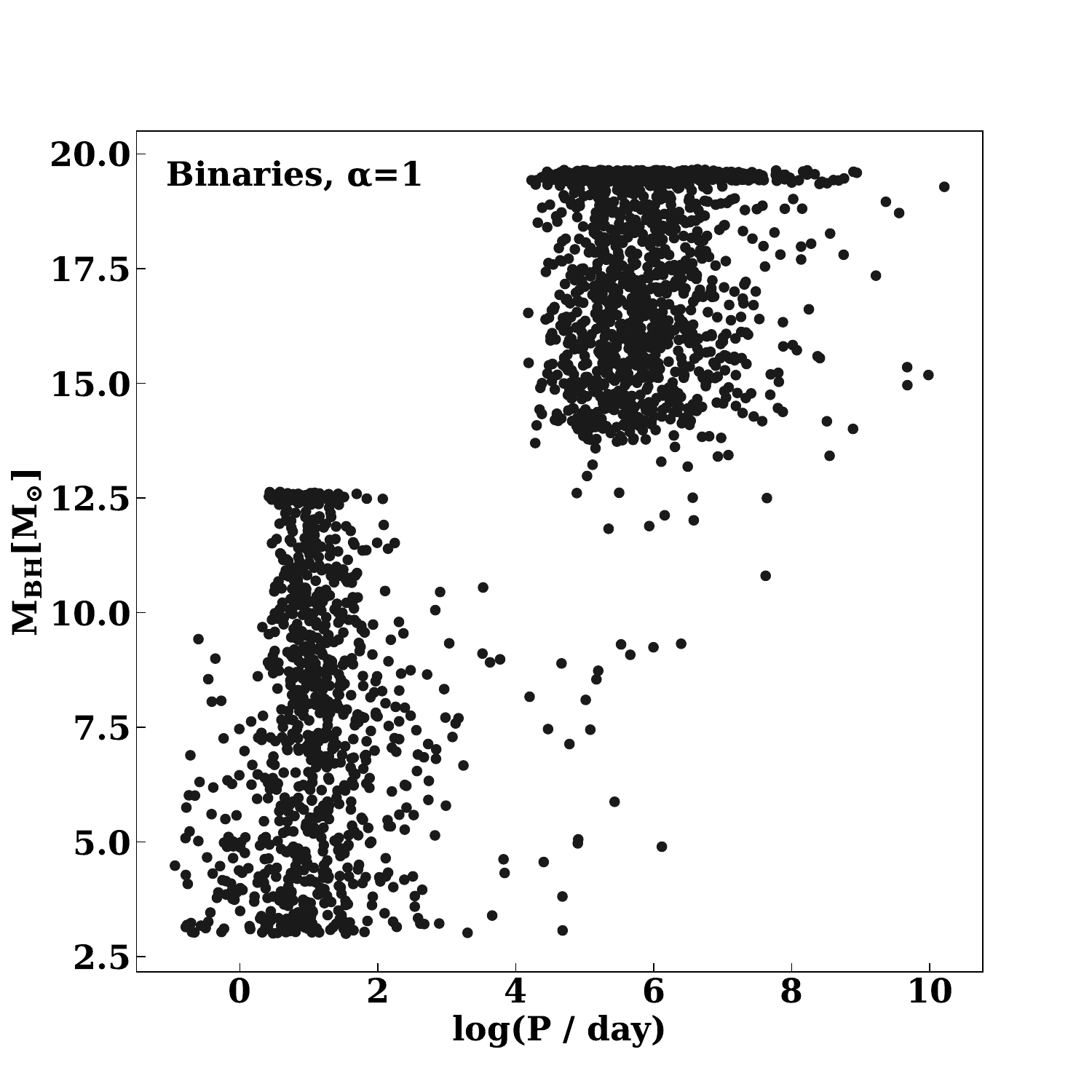}
    \includegraphics[width=\columnwidth]{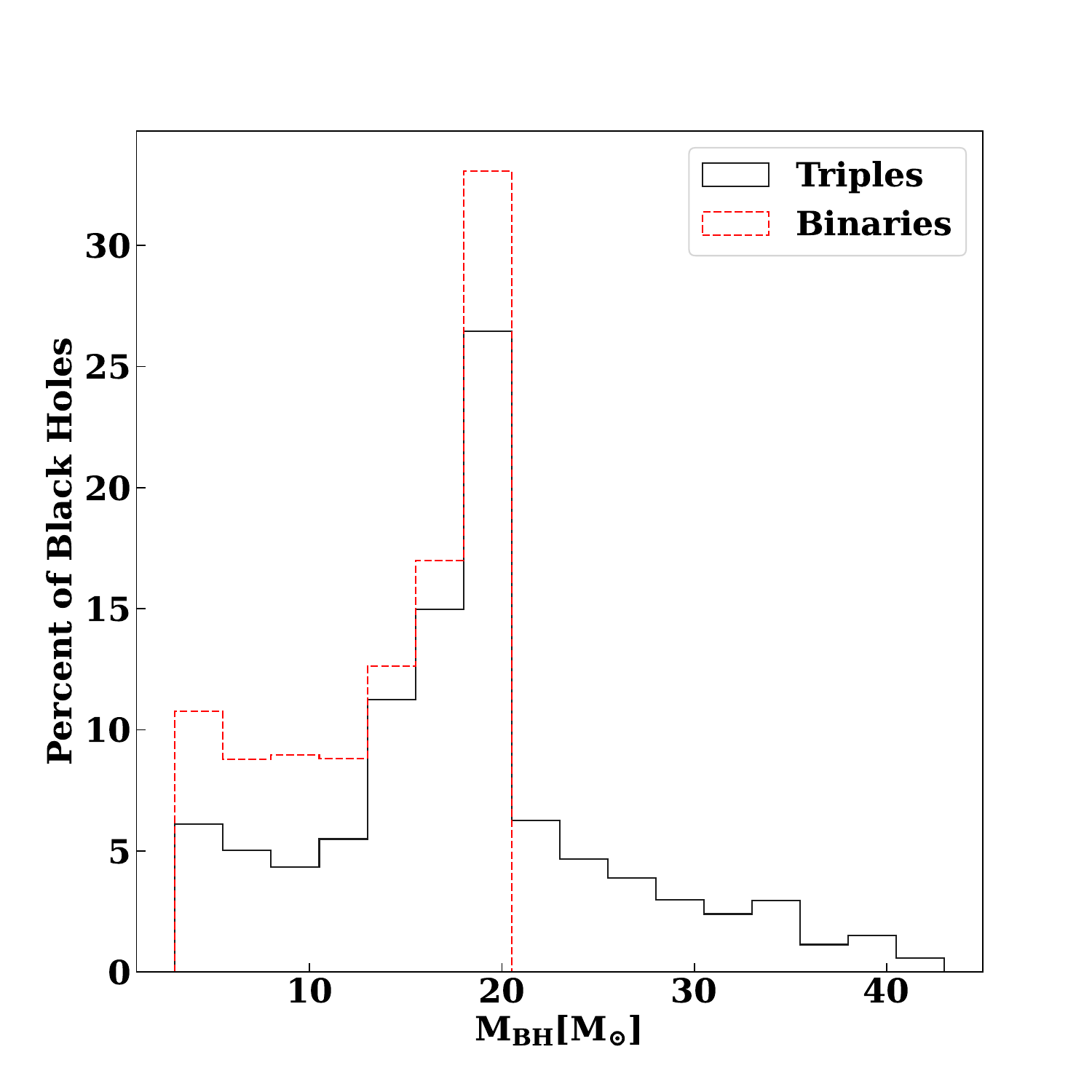}
    \figcaption{\label{fig:cloudPlot} Mass versus log period (in days) for BH-main sequence binaries from triples (top left) and binaries (top right) for $\alpha=1$. The bottom panel shows the BH mass distribution for each. Triples produce a more top-heavy BH mass function due to mergers in the inner binary.}
\end{figure*}

Mergers make the BH mass function more top-heavy. Figure~\ref{fig:cloudPlot} shows the BH mass as a function of period for BH-main sequence binaries originating in triples and binaries. The BH never exceeds $20 M_{\odot}$ for a binary initial condition but exceeds this threshold for $\sim 10-30\%$ of triple initial conditions. Almost all BH-main sequence with BHs above $20 M_{\odot}$ have periods $\geq 10^4$ days. 
Mergers that produce BHs above $20 M_{\odot}$ are not triggered by the tertiary companion, as they also occur when it is removed. However, in the binaries, the merger product is a single star and is thus not included in the distribution in Figure~\ref{fig:cloudPlot}.
 Additionally, the semi-major axis distribution affects the production of massive BHs. For the control binaries (where we do not apply the \citealt{mardling&aarseth2001} stability criterion), only $\sim .006\%$ of all BHs are above $20 M_{\odot}$, while for triples this fraction is 3\%.

\section{Rates}
\label{sec:rates}
We now estimate the formation efficiency of Gaia BH1-like systems for triples and binaries. We ran of order $10^6$ additional triple and of order $10^7$ additional binary simulations to reduce the uncertainty in our rate estimates. For efficiency, we require the secondary star to be between 0.5 and 2 $M_{\odot}$ in $\sim 1/2$ of the new triple simulations, considering the results in \S~\ref{sec:results}, where all Gaia BH1-like binaries form from the inner binary of the progenitor triple.

The fraction of Gaia BH1-like binaries in a stellar population, $f_{\rm GBH}$, is 

\begin{equation}
    f_{\rm GBH} = f_{\rm O} \times f_{\rm triple, binary} \times \underbrace{f_{\rm solar} \times f_{\rm eff}}_{\tilde{f}_{\rm eff}}.
    \label{eq:efficiency}
\end{equation}
Here $f_{O}$ is the fraction of BH progenitors (per stellar mass); $f_{\rm triple, binary}$ is the triple (or binary) fraction for these stars; $f_{\rm solar}$ is the fraction of massive triples (or binaries) with a companion between $0.5$ and $2 M_{\odot}$; $f_{\rm eff}$ is the fraction of such systems that produce Gaia BH1-like binaries. We group the last two terms into an overall efficiency, $\tilde{f}_{\rm eff}$.

For Kroupa mass function $f_{O}$ is $\approx 3.6\times 10^{-3} M_{\odot}^{-1}$ (assuming all stars above $18 M_{\odot}$ form BHs). Most massive stars are in triples or higher order multiples \citep{moe&distefano2017}. However, the exact breakdown by multiples is highly uncertain, especially for mass ratios $q<0.1$. We leave $f_{\rm triple, binary}$ unspecified, though $f_{\rm triple}$ is plausibly of order unity. $f_{\rm solar}\approx 0.26$ (0.12) for triples (binaries) based on the extrapolated \citet{moe&distefano2017} distributions.\footnote{$f_{\rm solar} \approx $ 0.093 for the inner binary of triples.} Finally, $f_{\rm eff}$ is estimated from the results in the previous section. 

The formation efficiencies are summarized in Table~\ref{tab:eff}, for triples and binaries. 
The efficiency is at least a factor of a few higher for triples than binaries.
However, for $\alpha=5$ 
the enhancement is mostly due to the larger fraction of solar-type stars in our triple initial conditions (the $f_{\rm solar}$ term in equation~\ref{eq:efficiency}). Otherwise, triples are not more efficient than binaries in producing Gaia BH1-like binaries for $\alpha=5$.
Overall, there are up to $2.9\times 10^{-7}$ Gaia BH1-like binaries per solar mass of stars formed.

Previous work has estimated the formation rate of BH-main sequence binaries in open star clusters \citep{rastello+2023, dicarlo+2023, tanikawa+2024}. For example, \citet{rastello+2023} find a formation efficiency of $\sim 2\times 10^{-7} M_{\odot}^{-1} M_{\odot}^{-1}$ for open clusters. This is within a factor of two of the maximum formation efficiency for triples. Other references give different rates (e.g. $2\times 10^{-6} M_{\odot}^{-1}$ in \citealp{tanikawa+2024}), but this is primarily due to difference definitions of Gaia BH1-like binaries.

We now estimate formation efficiency of all BH-MS binaries. Previously, \citet{dicarlo+2023} find a total of $4.2 \times 10^{-7}$ ($1.2\times 10^{-5}$) BH-MS binaries (with periods up to 10 yr\footnote{Wider systems would not be detectable by Gaia.}) per unit star formation from isolated binary evolution (open clusters). \citet{tanikawa+2024} obtained similar results.
To compare with these results, we ran additional binary and triple simulations with no restrictions on the companion masses. The formation efficiency BH-MS binaries (with periods less than 10 yr) in these simulations, is summarized in Table~\ref{tab:eff2}. 

The formation efficiency of BH-MS binaries for triples is enhanced by a factor of a few compared to binaries.\footnote{The formation efficiency for BH-MS from isolated evolution is an order of magnitude greater than in \citealt{dicarlo+2023}, mostly due to differences in the initial mass ratio distribution.} This enhancement is mostly due to the larger fraction of low mass companions in the triple initial conditions, compared to binaries. Overall, field triples are competitive with open clusters for forming BH-MS binaries.

\begin{deluxetable}{lccc}
        \tablecaption{\label{tab:eff} Formation efficiency of Gaia BH1-like binaries.
    }
         \tablehead{\colhead{Multiple-type}        & \colhead{$\alpha$}  &  \colhead{$\tilde{f}_{\rm eff}$\tablenotemark{a}} & \colhead{$f_{\rm GBH} / f_{\rm triple, binary}$\tablenotemark{b}}\\ \colhead{}                  & \colhead{}          &  \colhead{}                       & \colhead{$(M_{\odot}^{-1})$}}
\startdata
         Triple               &   1        &   $0.50-1.3\times 10^{-6}$    & $1.8-4.6\times 10^{-9}$\\
         Binary               &   1       &   $1 \times 10^{-8}$     &  $4 \times 10^{-11}$\\
         \hline
         Triple               &   5       &  $3.7-7.9 \times 10^{-5}$     & $1.3 - 2.9 \times 10^{-7}$\\
         Binary               &   5       &  $1.5-2.5 \times 10^{-5}$     & $5.2 - 9.0 \times 10^{-8}$\\
\enddata
\tablecomments{The Possonian uncertainties are  $\sim 10\%$, except for the second row. There the rate estimate is based on a single BH1-like system, and thus the uncertainty is of order unity.
\tablenotetext{a}{Gaia BH1-like binaries formed per massive triple or binary. The range shows the uncertainty from different definitions of Gaia BH1-like binaries. The smaller efficiency corresponds to the definition in \citet{rastello+2023}. For the larger efficiency, a Gaia BH1-like system is one where the eccentricity is within a factor of 1.5 of the observed system, while the BH mass, secondary mass, and period are within a factor of two of the observed system.}
\tablenotetext{b}{Gaia BH1-like binaries per $M_{\odot}$ of stars formed,  
    divided by the triple (or binary fraction) of massive stars.}}
\end{deluxetable}

\begin{deluxetable}{lccc}
\tablecaption{\label{tab:eff2} Formation efficiency of BH-main sequence (BH-MS) binaries with periods less than 10 yr.}
\tablehead{\colhead{Multiple-type}        & \colhead{$\alpha$}  &  \colhead{$\tilde{f}_{\rm eff}$\tablenotemark{a}} & \colhead{$f_{\rm BH-MS} / f_{\rm triple, binary}$\tablenotemark{b}}\\ \colhead{}                  & \colhead{}          &  \colhead{}                       & \colhead{$(M_{\odot}^{-1})$}}
\startdata
         Triple               &   1       &  $2.1 \times 10^{-3}$     &  $ 7.5 \times 10^{-6}$\\
         Binary               &   1       &  $5.5 \times 10^{-4}$     & $2.0 \times 10^{-6}$\\
         \hline
         Triple               &   5       &  $4.1 \times 10^{-3}$   & $1.5 \times 10^{-5}$\\
         Binary               &   5       &  $1.3 \times 10^{-3}$     & $4.7 \times 10^{-6}$\\
\enddata
\tablenotetext{a}{BH-MS binaries formed per massive triple or binary.}
\tablenotetext{b}{BH-MS binaries per $M_{\odot}$ of stars formed,  
    divided by the triple (or binary fraction) of massive stars.}
\end{deluxetable}

\subsection{Gaia BH2}
\citet{elbadry+2023bh2} reported the discovery of Gaia BH2: a BH-red giant system with a period of 1277 days and an eccentricity of 0.52. The mass of the BH is $8.94 \pm 0.34\, M_{\odot}$, while the companion is a $\sim 1 M_{\odot}$ red giant.  

For a common envelope efficiency, $\alpha=1$ ($\alpha=5$), our triple population synthesis produces 0 (3) Gaia BH2-like systems. These systems form from a merger in the inner binary followed by a common envelope. The primary and secondary then evolve into a BH and red giant.
This suggests $\lsim 7\times 10^{-9}$ Gaia BH2-like systems per star formed. 
Also, all of our Gaia BH2-like systems are more eccentric than the observed system and have eccentricities $\gsim$0.7. For isolated binaries, there are no Gaia BH2-like binaries for $\alpha=1$, while for $\alpha=5$ the formation efficiency is $\sim$an order of magnitude smaller than for triples.

Our models suggest that a large common envelope efficiency is required to reproduce Gaia BH2-like binaries, even for triples. However, the upper limit on the fraction of Gaia BH2-like systems per unit star formation is not very stringent due to small number statistics. (We have run $1.2 \times 10^6$ systems for $\alpha=1$, which gives an upper limit of $2.4\times 10^{-9} M_{\odot}^{-1}$ at 95\% confidence). Overall, more simulations are required to test the plausibility of producing Gaia BH2 through triples.

\section{Alternatives}
\label{sec:alt}
There are a few alternative scenarios for the observed Gaia BH binaries, besides those already discussed. We describe them here for completeness.

First, these systems could have formed from a binary with a very massive primary ($\gsim 50 M_{\odot}$) 
and a Sun-like secondary \citep{elbadry+2023bh2}. Such massive primaries may never become red supergiants \citep{humphreys&davidson1994,smith&conti2008,higgins&vink2019}. This behavior is not captured by our models. In this case, the observed periods are easily reproduced, considering Gaia BH1/BH2 need not have gone through a common envelope phase. 

Secondly, the interaction between the inner binary stars in a triple may have prevented either star from becoming a giant \citep{justham+2014}. This could lead to a low mass tertiary in orbit around a binary BH system \citep{elbadry+2023}. This behavior is also not captured by our models. In our triple synthesis, the only systems with binary BHs and a low mass tertiary have periods $\gsim 10^5$ days. Thus, they are much wider than the observed systems.

\section{Discussion and Summary}
\label{sec:conc}
Recently, a couple of candidate BH-stellar binaries have been identified in Gaia data: Gaia BH1 and Gaia BH2.
The former is a Sun-like main sequence star in a 185.5-day orbit around a 9.32 $M_{\odot}$ BH, while the 
latter is a red giant in a 1277-day orbit around an 8.94 $M_{\odot}$ BH.
These systems are difficult to explain with isolated binaries, potentially requiring unusual common envelope evolution \citep{elbadry+2023} and/or other scenarios, including formation in clusters.  

Here, we consider a triple scenario for the formation of such systems. Since, as known from observations, most BH progenitors are in triples or higher order multiples \citep{moe&distefano2017}, this is, to begin with, a more likely scenario than a binary progenitor.

Our main results are summarized as follows: 
\begin{enumerate}
\item Triple evolution allows for the formation of wider BH-main sequence systems at moderate eccentricities because secular eccentricity oscillations or instabilities allow systems to enter a common envelope at wider separations than in the binary case. 
\item As a result, systems with periods comparable to Gaia BH1 can be produced for smaller common envelope efficiencies (e.g. for $\alpha=1$), though there will only be a few$\times 10^{-9}$ per solar mass. For $\alpha=5$ both binaries and triples can produce Gaia BH1-like systems. The triple formation efficiency ($\sim 10^{-7} M_{\odot}^{-1}$)  is a factor of a few higher, due to the larger fraction of low mass companions in the triple initial conditions. For $\alpha=5$ the formation efficiency of Gaia BH1-like systems in triples is comparable to literature estimates for open clusters. Furthermore, triple dynamics in clusters may enhance the formation efficiency in these environments. We leave this to future study.
\item Gaia BH2 may require a larger common envelope efficiency, even in the triple scenario. However, more simulations are required to test this.
\item The formation efficiency for BH-main sequence binaries with periods $< 10$ yr is a factor of $3-4$ higher for triples than for binaries. Once again, this is most likely due to the larger fraction of low mass companions 
in our triple initial conditions, rather than differences in dynamics. The maximum formation efficiency of BH-MS systems in the triple channel is competitive with open clusters. 
\item BH-main sequence binaries from triples have a more top-heavy BH mass function than those from binaries, due to mergers within the inner binary. For example, the BH never exceeds $20 M_{\odot}$ for a binary initial condition but exceeds this threshold for $\sim 10-30\%$ of triple initial conditions. Almost all BH - main-sequence binaries with BH masses above $20 M_{\odot}$ have periods $\geq 10^4$ days. In isolation, such wide binaries would not give rise to any strong interactions, and would not form X-ray binaries, however, flyby encounters in the field can drive such binaries into interaction \citep{michaely&perets2016}. In clusters such wide binaries would be disrupted (soft binaries), but the BHs could be exchanged into other binaries, and even be part of BH-BH binaries that may merge through GW inspirals. Therefore, the elevated BH masses in these systems can affect the overall mass function of GW sources. 
In principle, flybys may also perturb triple phases of evolution \citep{michaely&perets2019, michaely&perets2020, shariat+2023}. However, in our case most triples are destroyed on timescales comparable to the lifetime of the massive primary.
\item The observed high triple and quadruple multiplicity among massive stars, and our results, among others, suggest that the use of binary population synthesis for the modeling of massive stars and of BHs may systematically provide incorrect evolutionary scenarios. 
\end{enumerate}

\section*{Acknowledgments}
We thank the anonymous referee for a constructive report. We thank Sara Rastello, Ataru Tanikawa, and Sukanya Chakrabarti for suggesting useful references. We thank Ugo Niccol\`{o} Di Carlo and Katelyn Breivik for providing information about their binary population synthesis calculations for the Gaia BH binaries. AG is supported at the Technion by a Zuckerman Fellowship.

\software{astropy \citep{astropy+2018}, Matplotlib \citep{hunter+2007}, NumPy, SciPy \citep{2020SciPy-NMeth},
MSE \citep{hamers+2021},
COSMIC \citep{breivik+2020}}


\bibliographystyle{aasjournal}
\bibliography{main} 

\begin{thebibliography}{}
\expandafter\ifx\csname natexlab\endcsname\relax\def\natexlab#1{#1}\fi
\providecommand{\url}[1]{\href{#1}{#1}}
\providecommand{\dodoi}[1]{doi:~\href{http://doi.org/#1}{\nolinkurl{#1}}}
\providecommand{\doeprint}[1]{\href{http://ascl.net/#1}{\nolinkurl{http://ascl.net/#1}}}
\providecommand{\doarXiv}[1]{\href{https://arxiv.org/abs/#1}{\nolinkurl{https://arxiv.org/abs/#1}}}

\bibitem[{{Andrews} {et~al.}(2022){Andrews}, {Taggart}, \&
  {Foley}}]{andrews+2022}
{Andrews}, J.~J., {Taggart}, K., \& {Foley}, R. 2022, arXiv e-prints,
  arXiv:2207.00680, \dodoi{10.48550/arXiv.2207.00680}

\bibitem[{{Antognini} {et~al.}(2014){Antognini}, {Shappee}, {Thompson}, \&
  {Amaro-Seoane}}]{Ant+14}
{Antognini}, J.~M., {Shappee}, B.~J., {Thompson}, T.~A., \& {Amaro-Seoane}, P.
  2014, \mnras, 439, 1079, \dodoi{10.1093/mnras/stu039}

\bibitem[{{Antognini}(2015)}]{antognini2015}
{Antognini}, J.~M.~O. 2015, \mnras, 452, 3610, \dodoi{10.1093/mnras/stv1552}

\bibitem[{{Antonini} {et~al.}(2017){Antonini}, {Toonen}, \& {Hamers}}]{Ant+17}
{Antonini}, F., {Toonen}, S., \& {Hamers}, A.~S. 2017, \apj, 841, 77,
  \dodoi{10.3847/1538-4357/aa6f5e}

\bibitem[{{Breivik} {et~al.}(2020){Breivik}, {Coughlin}, {Zevin}, {Rodriguez},
  {Kremer}, {Ye}, {Andrews}, {Kurkowski}, {Digman}, {Larson}, \&
  {Rasio}}]{breivik+2020}
{Breivik}, K., {Coughlin}, S., {Zevin}, M., {et~al.} 2020, \apj, 898, 71,
  \dodoi{10.3847/1538-4357/ab9d85}

\bibitem[{{Chakrabarti} {et~al.}(2023){Chakrabarti}, {Simon}, {Craig},
  {Reggiani}, {Brandt}, {Guhathakurta}, {Dalba}, {Kirby}, {Chang}, {Hey},
  {Savino}, {Geha}, \& {Thompson}}]{chakrabarti+2023}
{Chakrabarti}, S., {Simon}, J.~D., {Craig}, P.~A., {et~al.} 2023, \aj, 166, 6,
  \dodoi{10.3847/1538-3881/accf21}

\bibitem[{{Corral-Santana} {et~al.}(2016){Corral-Santana}, {Casares},
  {Mu{\~n}oz-Darias}, {Bauer}, {Mart{\'\i}nez-Pais}, \&
  {Russell}}]{corral-santana+2016}
{Corral-Santana}, J.~M., {Casares}, J., {Mu{\~n}oz-Darias}, T., {et~al.} 2016,
  \aap, 587, A61, \dodoi{10.1051/0004-6361/201527130}

\bibitem[{{Di Carlo} {et~al.}(2023){Di Carlo}, {Agrawal}, {Rodriguez}, \&
  {Breivik}}]{dicarlo+2023}
{Di Carlo}, U.~N., {Agrawal}, P., {Rodriguez}, C.~L., \& {Breivik}, K. 2023,
  arXiv e-prints, arXiv:2306.13121, \dodoi{10.48550/arXiv.2306.13121}

\bibitem[{{El-Badry} {et~al.}(2023{\natexlab{a}}){El-Badry}, {Rix}, {Quataert},
  {Howard}, {Isaacson}, {Fuller}, {Hawkins}, {Breivik}, {Wong}, {Rodriguez},
  {Conroy}, {Shahaf}, {Mazeh}, {Arenou}, {Burdge}, {Bashi}, {Faigler}, {Weisz},
  {Seeburger}, {Almada Monter}, \& {Wojno}}]{elbadry+2023}
{El-Badry}, K., {Rix}, H.-W., {Quataert}, E., {et~al.} 2023{\natexlab{a}},
  \mnras, 518, 1057, \dodoi{10.1093/mnras/stac3140}

\bibitem[{{El-Badry} {et~al.}(2023{\natexlab{b}}){El-Badry}, {Rix}, {Cendes},
  {Rodriguez}, {Conroy}, {Quataert}, {Hawkins}, {Zari}, {Hobson}, {Breivik},
  {Rau}, {Berger}, {Shahaf}, {Seeburger}, {Burdge}, {Latham}, {Buchhave},
  {Bieryla}, {Bashi}, {Mazeh}, \& {Faigler}}]{elbadry+2023bh2}
{El-Badry}, K., {Rix}, H.-W., {Cendes}, Y., {et~al.} 2023{\natexlab{b}},
  \mnras, 521, 4323, \dodoi{10.1093/mnras/stad799}

\bibitem[{{Fortin} {et~al.}(2023){Fortin}, {Garc{\'\i}a}, {Simaz Bunzel}, \&
  {Chaty}}]{fortin+2023}
{Fortin}, F., {Garc{\'\i}a}, F., {Simaz Bunzel}, A., \& {Chaty}, S. 2023, \aap,
  671, A149, \dodoi{10.1051/0004-6361/202245236}

\bibitem[{{Fryer} {et~al.}(2012){Fryer}, {Belczynski}, {Wiktorowicz},
  {Dominik}, {Kalogera}, \& {Holz}}]{fryer+2012}
{Fryer}, C.~L., {Belczynski}, K., {Wiktorowicz}, G., {et~al.} 2012, \apj, 749,
  91, \dodoi{10.1088/0004-637X/749/1/91}

\bibitem[{{Gaia Collaboration} {et~al.}(2023{\natexlab{a}}){Gaia
  Collaboration}, {Vallenari}, {Brown}, {Prusti}, {de Bruijne}, {Arenou},
  {Babusiaux}, {Biermann}, {Creevey}, {Ducourant}, {Evans}, {Eyer}, {Guerra},
  {Hutton}, {Jordi}, {Klioner}, {Lammers}, {Lindegren}, {Luri}, {Mignard},
  {Panem}, {Pourbaix}, {Randich}, {Sartoretti}, {Soubiran}, {Tanga}, {Walton},
  {Bailer-Jones}, {Bastian}, {Drimmel}, {Jansen}, {Katz}, {Lattanzi}, {van
  Leeuwen}, {Bakker}, {Cacciari}, {Casta{\~n}eda}, {De Angeli}, {Fabricius},
  {Fouesneau}, {Fr{\'e}mat}, {Galluccio}, {Guerrier}, {Heiter}, {Masana},
  {Messineo}, {Mowlavi}, {Nicolas}, {Nienartowicz}, {Pailler}, {Panuzzo},
  {Riclet}, {Roux}, {Seabroke}, {Sordo}, {Th{\'e}venin}, {Gracia-Abril},
  {Portell}, {Teyssier}, {Altmann}, {Andrae}, {Audard}, {Bellas-Velidis},
  {Benson}, {Berthier}, {Blomme}, {Burgess}, {Busonero}, {Busso},
  {C{\'a}novas}, {Carry}, {Cellino}, {Cheek}, {Clementini}, {Damerdji},
  {Davidson}, {de Teodoro}, {Nu{\~n}ez Campos}, {Delchambre}, {Dell'Oro},
  {Esquej}, {Fern{\'a}ndez-Hern{\'a}ndez}, {Fraile}, {Garabato},
  {Garc{\'\i}a-Lario}, {Gosset}, {Haigron}, {Halbwachs}, {Hambly}, {Harrison},
  {Hern{\'a}ndez}, {Hestroffer}, {Hodgkin}, {Holl}, {Jan{\ss}en}, {Jevardat de
  Fombelle}, {Jordan}, {Krone-Martins}, {Lanzafame}, {L{\"o}ffler}, {Marchal},
  {Marrese}, {Moitinho}, {Muinonen}, {Osborne}, {Pancino}, {Pauwels},
  {Recio-Blanco}, {Reyl{\'e}}, {Riello}, {Rimoldini}, {Roegiers}, {Rybizki},
  {Sarro}, {Siopis}, {Smith}, {Sozzetti}, {Utrilla}, {van Leeuwen}, {Abbas},
  {{\'A}brah{\'a}m}, {Abreu Aramburu}, {Aerts}, {Aguado}, {Ajaj},
  {Aldea-Montero}, {Altavilla}, {{\'A}lvarez}, {Alves}, {Anders}, {Anderson},
  {Anglada Varela}, {Antoja}, {Baines}, {Baker}, {Balaguer-N{\'u}{\~n}ez},
  {Balbinot}, {Balog}, {Barache}, {Barbato}, {Barros}, {Barstow},
  {Bartolom{\'e}}, {Bassilana}, {Bauchet}, {Becciani}, {Bellazzini},
  {Berihuete}, {Bernet}, {Bertone}, {Bianchi}, {Binnenfeld}, {Blanco-Cuaresma},
  {Blazere}, {Boch}, {Bombrun}, {Bossini}, {Bouquillon}, {Bragaglia},
  {Bramante}, {Breedt}, {Bressan}, {Brouillet}, {Brugaletta}, {Bucciarelli},
  {Burlacu}, {Butkevich}, {Buzzi}, {Caffau}, {Cancelliere}, {Cantat-Gaudin},
  {Carballo}, {Carlucci}, {Carnerero}, {Carrasco}, {Casamiquela}, {Castellani},
  {Castro-Ginard}, {Chaoul}, {Charlot}, {Chemin}, {Chiaramida}, {Chiavassa},
  {Chornay}, {Comoretto}, {Contursi}, {Cooper}, {Cornez}, {Cowell}, {Crifo},
  {Cropper}, {Crosta}, {Crowley}, {Dafonte}, {Dapergolas}, {David}, {David},
  {de Laverny}, {De Luise}, {De March}, {De Ridder}, {de Souza}, {de Torres},
  {del Peloso}, {del Pozo}, {Delbo}, {Delgado}, {Delisle}, {Demouchy},
  {Dharmawardena}, {Di Matteo}, {Diakite}, {Diener}, {Distefano}, {Dolding},
  {Edvardsson}, {Enke}, {Fabre}, {Fabrizio}, {Faigler}, {Fedorets}, {Fernique},
  {Fienga}, {Figueras}, {Fournier}, {Fouron}, {Fragkoudi}, {Gai},
  {Garcia-Gutierrez}, {Garcia-Reinaldos}, {Garc{\'\i}a-Torres}, {Garofalo},
  {Gavel}, {Gavras}, {Gerlach}, {Geyer}, {Giacobbe}, {Gilmore}, {Girona},
  {Giuffrida}, {Gomel}, {Gomez}, {Gonz{\'a}lez-N{\'u}{\~n}ez},
  {Gonz{\'a}lez-Santamar{\'\i}a}, {Gonz{\'a}lez-Vidal}, {Granvik}, {Guillout},
  {Guiraud}, {Guti{\'e}rrez-S{\'a}nchez}, {Guy}, {Hatzidimitriou}, {Hauser},
  {Haywood}, {Helmer}, {Helmi}, {Sarmiento}, {Hidalgo}, {Hilger},
  {H{\l}adczuk}, {Hobbs}, {Holland}, {Huckle}, {Jardine}, {Jasniewicz},
  {Jean-Antoine Piccolo}, {Jim{\'e}nez-Arranz}, {Jorissen}, {Juaristi
  Campillo}, {Julbe}, {Karbevska}, {Kervella}, {Khanna}, {Kontizas},
  {Kordopatis}, {Korn}, {K{\'o}sp{\'a}l}, {Kostrzewa-Rutkowska},
  {Kruszy{\'n}ska}, {Kun}, {Laizeau}, {Lambert}, {Lanza}, {Lasne}, {Le
  Campion}, {Lebreton}, {Lebzelter}, {Leccia}, {Leclerc}, {Lecoeur-Taibi},
  {Liao}, {Licata}, {Lindstr{\o}m}, {Lister}, {Livanou}, {Lobel}, {Lorca},
  {Loup}, {Madrero Pardo}, {Magdaleno Romeo}, {Managau}, {Mann}, {Manteiga},
  {Marchant}, {Marconi}, {Marcos}, {Marcos Santos}, {Mar{\'\i}n Pina},
  {Marinoni}, {Marocco}, {Marshall}, {Martin Polo}, {Mart{\'\i}n-Fleitas},
  {Marton}, {Mary}, {Masip}, {Massari}, {Mastrobuono-Battisti}, {Mazeh},
  {McMillan}, {Messina}, {Michalik}, {Millar}, {Mints}, {Molina}, {Molinaro},
  {Moln{\'a}r}, {Monari}, {Mongui{\'o}}, {Montegriffo}, {Montero}, {Mor},
  {Mora}, {Morbidelli}, {Morel}, {Morris}, {Muraveva}, {Murphy}, {Musella},
  {Nagy}, {Noval}, {Oca{\~n}a}, {Ogden}, {Ordenovic}, {Osinde}, {Pagani},
  {Pagano}, {Palaversa}, {Palicio}, {Pallas-Quintela}, {Panahi},
  {Payne-Wardenaar}, {Pe{\~n}alosa Esteller}, {Penttil{\"a}}, {Pichon},
  {Piersimoni}, {Pineau}, {Plachy}, {Plum}, {Poggio}, {Pr{\v{s}}a}, {Pulone},
  {Racero}, {Ragaini}, {Rainer}, {Raiteri}, {Rambaux}, {Ramos}, {Ramos-Lerate},
  {Re Fiorentin}, {Regibo}, {Richards}, {Rios Diaz}, {Ripepi}, {Riva}, {Rix},
  {Rixon}, {Robichon}, {Robin}, {Robin}, {Roelens}, {Rogues}, {Rohrbasser},
  {Romero-G{\'o}mez}, {Rowell}, {Royer}, {Ruz Mieres}, {Rybicki}, {Sadowski},
  {S{\'a}ez N{\'u}{\~n}ez}, {Sagrist{\`a} Sell{\'e}s}, {Sahlmann}, {Salguero},
  {Samaras}, {Sanchez Gimenez}, {Sanna}, {Santove{\~n}a}, {Sarasso},
  {Schultheis}, {Sciacca}, {Segol}, {Segovia}, {S{\'e}gransan}, {Semeux},
  {Shahaf}, {Siddiqui}, {Siebert}, {Siltala}, {Silvelo}, {Slezak}, {Slezak},
  {Smart}, {Snaith}, {Solano}, {Solitro}, {Souami}, {Souchay}, {Spagna},
  {Spina}, {Spoto}, {Steele}, {Steidelm{\"u}ller}, {Stephenson}, {S{\"u}veges},
  {Surdej}, {Szabados}, {Szegedi-Elek}, {Taris}, {Taylor}, {Teixeira},
  {Tolomei}, {Tonello}, {Torra}, {Torra}, {Torralba Elipe}, {Trabucchi},
  {Tsounis}, {Turon}, {Ulla}, {Unger}, {Vaillant}, {van Dillen}, {van Reeven},
  {Vanel}, {Vecchiato}, {Viala}, {Vicente}, {Voutsinas}, {Weiler}, {Wevers},
  {Wyrzykowski}, {Yoldas}, {Yvard}, {Zhao}, {Zorec}, {Zucker}, \&
  {Zwitter}}]{gaia+2023a}
{Gaia Collaboration}, {Vallenari}, A., {Brown}, A.~G.~A., {et~al.}
  2023{\natexlab{a}}, \aap, 674, A1, \dodoi{10.1051/0004-6361/202243940}

\bibitem[{{Gaia Collaboration} {et~al.}(2023{\natexlab{b}}){Gaia
  Collaboration}, {Arenou}, {Babusiaux}, {Barstow}, {Faigler}, {Jorissen},
  {Kervella}, {Mazeh}, {Mowlavi}, {Panuzzo}, {Sahlmann}, {Shahaf}, {Sozzetti},
  {Bauchet}, {Damerdji}, {Gavras}, {Giacobbe}, {Gosset}, {Halbwachs}, {Holl},
  {Lattanzi}, {Leclerc}, {Morel}, {Pourbaix}, {Re Fiorentin}, {Sadowski},
  {S{\'e}gransan}, {Siopis}, {Teyssier}, {Zwitter}, {Planquart}, {Brown},
  {Vallenari}, {Prusti}, {de Bruijne}, {Biermann}, {Creevey}, {Ducourant},
  {Evans}, {Eyer}, {Guerra}, {Hutton}, {Jordi}, {Klioner}, {Lammers},
  {Lindegren}, {Luri}, {Mignard}, {Panem}, {Randich}, {Sartoretti}, {Soubiran},
  {Tanga}, {Walton}, {Bailer-Jones}, {Bastian}, {Drimmel}, {Jansen}, {Katz},
  {van Leeuwen}, {Bakker}, {Cacciari}, {Casta{\~n}eda}, {De Angeli},
  {Fabricius}, {Fouesneau}, {Fr{\'e}mat}, {Galluccio}, {Guerrier}, {Heiter},
  {Masana}, {Messineo}, {Nicolas}, {Nienartowicz}, {Pailler}, {Riclet}, {Roux},
  {Seabroke}, {Sordo}, {Th{\'e}venin}, {Gracia-Abril}, {Portell}, {Altmann},
  {Andrae}, {Audard}, {Bellas-Velidis}, {Benson}, {Berthier}, {Blomme},
  {Burgess}, {Busonero}, {Busso}, {C{\'a}novas}, {Carry}, {Cellino}, {Cheek},
  {Clementini}, {Davidson}, {de Teodoro}, {Nu{\~n}ez Campos}, {Delchambre},
  {Dell'Oro}, {Esquej}, {Fern{\'a}ndez-Hern{\'a}ndez}, {Fraile}, {Garabato},
  {Garc{\'\i}a-Lario}, {Haigron}, {Hambly}, {Harrison}, {Hern{\'a}ndez},
  {Hestroffer}, {Hodgkin}, {Jan{\ss}en}, {Jevardat de Fombelle}, {Jordan},
  {Krone-Martins}, {Lanzafame}, {L{\"o}ffler}, {Marchal}, {Marrese},
  {Moitinho}, {Muinonen}, {Osborne}, {Pancino}, {Pauwels}, {Recio-Blanco},
  {Reyl{\'e}}, {Riello}, {Rimoldini}, {Roegiers}, {Rybizki}, {Sarro}, {Smith},
  {Utrilla}, {van Leeuwen}, {Abbas}, {{\'A}brah{\'a}m}, {Abreu Aramburu},
  {Aerts}, {Aguado}, {Ajaj}, {Aldea-Montero}, {Altavilla}, {{\'A}lvarez},
  {Alves}, {Anders}, {Anderson}, {Anglada Varela}, {Antoja}, {Baines}, {Baker},
  {Balaguer-N{\'u}{\~n}ez}, {Balbinot}, {Balog}, {Barache}, {Barbato},
  {Barros}, {Bartolom{\'e}}, {Bassilana}, {Becciani}, {Bellazzini},
  {Berihuete}, {Bernet}, {Bertone}, {Bianchi}, {Binnenfeld}, {Blanco-Cuaresma},
  {Blazere}, {Boch}, {Bombrun}, {Bossini}, {Bouquillon}, {Bragaglia},
  {Bramante}, {Breedt}, {Bressan}, {Brouillet}, {Brugaletta}, {Bucciarelli},
  {Burlacu}, {Butkevich}, {Buzzi}, {Caffau}, {Cancelliere}, {Cantat-Gaudin},
  {Carballo}, {Carlucci}, {Carnerero}, {Carrasco}, {Casamiquela}, {Castellani},
  {Castro-Ginard}, {Chaoul}, {Charlot}, {Chemin}, {Chiaramida}, {Chiavassa},
  {Chornay}, {Comoretto}, {Contursi}, {Cooper}, {Cornez}, {Cowell}, {Crifo},
  {Cropper}, {Crosta}, {Crowley}, {Dafonte}, {Dapergolas}, {David}, {de
  Laverny}, {De Luise}, {De March}, {De Ridder}, {de Souza}, {de Torres}, {del
  Peloso}, {del Pozo}, {Delbo}, {Delgado}, {Delisle}, {Demouchy},
  {Dharmawardena}, {Diakite}, {Diener}, {Distefano}, {Dolding}, {Enke},
  {Fabre}, {Fabrizio}, {Fedorets}, {Fernique}, {Figueras}, {Fournier},
  {Fouron}, {Fragkoudi}, {Gai}, {Garcia-Gutierrez}, {Garcia-Reinaldos},
  {Garc{\'\i}a-Torres}, {Garofalo}, {Gavel}, {Gerlach}, {Geyer}, {Gilmore},
  {Girona}, {Giuffrida}, {Gomel}, {Gomez}, {Gonz{\'a}lez-N{\'u}{\~n}ez},
  {Gonz{\'a}lez-Santamar{\'\i}a}, {Gonz{\'a}lez-Vidal}, {Granvik}, {Guillout},
  {Guiraud}, {Guti{\'e}rrez-S{\'a}nchez}, {Guy}, {Hatzidimitriou}, {Hauser},
  {Haywood}, {Helmer}, {Helmi}, {Sarmiento}, {Hidalgo}, {Hilger},
  {H{\l}adczuk}, {Hobbs}, {Holland}, {Huckle}, {Jardine}, {Jasniewicz},
  {Jean-Antoine Piccolo}, {Jim{\'e}nez-Arranz}, {Juaristi Campillo}, {Julbe},
  {Karbevska}, {Khanna}, {Kordopatis}, {Korn}, {K{\'o}sp{\'a}l},
  {Kostrzewa-Rutkowska}, {Kruszy{\'n}ska}, {Kun}, {Laizeau}, {Lambert},
  {Lanza}, {Lasne}, {Le Campion}, {Lebreton}, {Lebzelter}, {Leccia},
  {Lecoeur-Taibi}, {Liao}, {Licata}, {Lindstr{\o}m}, {Lister}, {Livanou},
  {Lobel}, {Lorca}, {Loup}, {Madrero Pardo}, {Magdaleno Romeo}, {Managau},
  {Mann}, {Manteiga}, {Marchant}, {Marconi}, {Marcos}, {Marcos Santos},
  {Mar{\'\i}n Pina}, {Marinoni}, {Marocco}, {Marshall}, {Martin Polo},
  {Mart{\'\i}n-Fleitas}, {Marton}, {Mary}, {Masip}, {Massari},
  {Mastrobuono-Battisti}, {McMillan}, {Messina}, {Michalik}, {Millar}, {Mints},
  {Molina}, {Molinaro}, {Moln{\'a}r}, {Monari}, {Mongui{\'o}}, {Montegriffo},
  {Montero}, {Mor}, {Mora}, {Morbidelli}, {Morris}, {Muraveva}, {Murphy},
  {Musella}, {Nagy}, {Noval}, {Oca{\~n}a}, {Ogden}, {Ordenovic}, {Osinde},
  {Pagani}, {Pagano}, {Palaversa}, {Palicio}, {Pallas-Quintela}, {Panahi},
  {Payne-Wardenaar}, {Pe{\~n}alosa Esteller}, {Penttil{\"a}}, {Pichon},
  {Piersimoni}, {Pineau}, {Plachy}, {Plum}, {Poggio}, {Pr{\v{s}}a}, {Pulone},
  {Racero}, {Ragaini}, {Rainer}, {Raiteri}, {Ramos}, {Ramos-Lerate}, {Regibo},
  {Richards}, {Rios Diaz}, {Ripepi}, {Riva}, {Rix}, {Rixon}, {Robichon},
  {Robin}, {Robin}, {Roelens}, {Rogues}, {Rohrbasser}, {Romero-G{\'o}mez},
  {Rowell}, {Royer}, {Ruz Mieres}, {Rybicki}, {S{\'a}ez N{\'u}{\~n}ez},
  {Sagrist{\`a} Sell{\'e}s}, {Salguero}, {Samaras}, {Sanchez Gimenez}, {Sanna},
  {Santove{\~n}a}, {Sarasso}, {Schultheis}, {Sciacca}, {Segol}, {Segovia},
  {Semeux}, {Siddiqui}, {Siebert}, {Siltala}, {Silvelo}, {Slezak}, {Slezak},
  {Smart}, {Snaith}, {Solano}, {Solitro}, {Souami}, {Souchay}, {Spagna},
  {Spina}, {Spoto}, {Steele}, {Steidelm{\"u}ller}, {Stephenson}, {S{\"u}veges},
  {Surdej}, {Szabados}, {Szegedi-Elek}, {Taris}, {Taylor}, {Teixeira},
  {Tolomei}, {Tonello}, {Torra}, {Torra}, {Torralba Elipe}, {Trabucchi},
  {Tsounis}, {Turon}, {Ulla}, {Unger}, {Vaillant}, {van Dillen}, {van Reeven},
  {Vanel}, {Vecchiato}, {Viala}, {Vicente}, {Voutsinas}, {Weiler}, {Wevers},
  {Wyrzykowski}, {Yoldas}, {Yvard}, {Zhao}, {Zorec}, \& {Zucker}}]{gaia+2023b}
{Gaia Collaboration}, {Arenou}, F., {Babusiaux}, C., {et~al.}
  2023{\natexlab{b}}, \aap, 674, A34, \dodoi{10.1051/0004-6361/202243782}

\bibitem[{{Hamers} \& {Dosopoulou}(2019)}]{hamers&dosopoulou2019}
{Hamers}, A.~S., \& {Dosopoulou}, F. 2019, \apj, 872, 119,
  \dodoi{10.3847/1538-4357/ab001d}

\bibitem[{{Hamers} {et~al.}(2022){Hamers}, {Perets}, {Thompson}, \&
  {Neunteufel}}]{hamers+2022}
{Hamers}, A.~S., {Perets}, H.~B., {Thompson}, T.~A., \& {Neunteufel}, P. 2022,
  \apj, 925, 178, \dodoi{10.3847/1538-4357/ac400b}

\bibitem[{{Hamers} {et~al.}(2021){Hamers}, {Rantala}, {Neunteufel}, {Preece},
  \& {Vynatheya}}]{hamers+2021}
{Hamers}, A.~S., {Rantala}, A., {Neunteufel}, P., {Preece}, H., \& {Vynatheya},
  P. 2021, \mnras, 502, 4479, \dodoi{10.1093/mnras/stab287}

\bibitem[{{Higgins} \& {Vink}(2019)}]{higgins&vink2019}
{Higgins}, E.~R., \& {Vink}, J.~S. 2019, \aap, 622, A50,
  \dodoi{10.1051/0004-6361/201834123}

\bibitem[{{Humphreys} \& {Davidson}(1994)}]{humphreys&davidson1994}
{Humphreys}, R.~M., \& {Davidson}, K. 1994, \pasp, 106, 1025,
  \dodoi{10.1086/133478}

\bibitem[{Hunter(2007)}]{hunter+2007}
Hunter, J.~D. 2007, Computing in Science \& Engineering, 9, 90,
  \dodoi{10.1109/MCSE.2007.55}

\bibitem[{{Hurley} {et~al.}(2002){Hurley}, {Tout}, \& {Pols}}]{hurley+2002}
{Hurley}, J.~R., {Tout}, C.~A., \& {Pols}, O.~R. 2002, \mnras, 329, 897,
  \dodoi{10.1046/j.1365-8711.2002.05038.x}

\bibitem[{{Justham} {et~al.}(2014){Justham}, {Podsiadlowski}, \&
  {Vink}}]{justham+2014}
{Justham}, S., {Podsiadlowski}, P., \& {Vink}, J.~S. 2014, \apj, 796, 121,
  \dodoi{10.1088/0004-637X/796/2/121}

\bibitem[{{Kozai}(1962)}]{kozai1962}
{Kozai}, Y. 1962, \aj, 67, 591, \dodoi{10.1086/108790}

\bibitem[{{Kummer} {et~al.}(2023){Kummer}, {Toonen}, \& {de
  Koter}}]{kummer+2023}
{Kummer}, F., {Toonen}, S., \& {de Koter}, A. 2023, \aap, 678, A60,
  \dodoi{10.1051/0004-6361/202347179}

\bibitem[{{Lam} {et~al.}(2022){Lam}, {Lu}, {Udalski}, {Bond}, {Bennett},
  {Skowron}, {Mr{\'o}z}, {Poleski}, {Sumi}, {Szyma{\'n}ski}, {Koz{\l}owski},
  {Pietrukowicz}, {Soszy{\'n}ski}, {Ulaczyk}, {Wyrzykowski}, {Miyazaki},
  {Suzuki}, {Koshimoto}, {Rattenbury}, {Hosek}, {Abe}, {Barry}, {Bhattacharya},
  {Fukui}, {Fujii}, {Hirao}, {Itow}, {Kirikawa}, {Kondo}, {Matsubara},
  {Matsumoto}, {Muraki}, {Olmschenk}, {Ranc}, {Okamura}, {Satoh}, {Silva},
  {Toda}, {Tristram}, {Vandorou}, {Yama}, {Abrams}, {Agarwal}, {Rose}, \&
  {Terry}}]{lam+2022}
{Lam}, C.~Y., {Lu}, J.~R., {Udalski}, A., {et~al.} 2022, \apjl, 933, L23,
  \dodoi{10.3847/2041-8213/ac7442}

\bibitem[{{Lidov}(1962)}]{lidov1962}
{Lidov}, M.~L. 1962, \planss, 9, 719, \dodoi{10.1016/0032-0633(62)90129-0}

\bibitem[{{Mardling} \& {Aarseth}(2001)}]{mardling&aarseth2001}
{Mardling}, R.~A., \& {Aarseth}, S.~J. 2001, \mnras, 321, 398,
  \dodoi{10.1046/j.1365-8711.2001.03974.x}

\bibitem[{{Michaely} \& {Perets}(2014)}]{Mic+14}
{Michaely}, E., \& {Perets}, H.~B. 2014, \apj, 794, 122,
  \dodoi{10.1088/0004-637X/794/2/122}

\bibitem[{{Michaely} \& {Perets}(2016)}]{michaely&perets2016}
---. 2016, \mnras, 458, 4188, \dodoi{10.1093/mnras/stw368}

\bibitem[{{Michaely} \& {Perets}(2019)}]{michaely&perets2019}
---. 2019, \apjl, 887, L36, \dodoi{10.3847/2041-8213/ab5b9b}

\bibitem[{{Michaely} \& {Perets}(2020)}]{michaely&perets2020}
---. 2020, \mnras, 498, 4924, \dodoi{10.1093/mnras/staa2720}

\bibitem[{{Moe} \& {Di Stefano}(2017)}]{moe&distefano2017}
{Moe}, M., \& {Di Stefano}, R. 2017, \apjs, 230, 15,
  \dodoi{10.3847/1538-4365/aa6fb6}

\bibitem[{{Mr{\'o}z} {et~al.}(2022){Mr{\'o}z}, {Udalski}, \&
  {Gould}}]{mroz+2022}
{Mr{\'o}z}, P., {Udalski}, A., \& {Gould}, A. 2022, \apjl, 937, L24,
  \dodoi{10.3847/2041-8213/ac90bb}

\bibitem[{{Naoz}(2016)}]{naoz+2016}
{Naoz}, S. 2016, \araa, 54, 441, \dodoi{10.1146/annurev-astro-081915-023315}

\bibitem[{{Perets} \& {Fabrycky}(2009)}]{perets&fabrycky2009}
{Perets}, H.~B., \& {Fabrycky}, D.~C. 2009, \apj, 697, 1048,
  \dodoi{10.1088/0004-637X/697/2/1048}

\bibitem[{{Perets} \& {Kratter}(2012)}]{perets&kratter2012}
{Perets}, H.~B., \& {Kratter}, K.~M. 2012, \apj, 760, 99,
  \dodoi{10.1088/0004-637X/760/2/99}

\bibitem[{{Rastello} {et~al.}(2023){Rastello}, {Iorio}, {Mapelli},
  {Arca-Sedda}, {Di Carlo}, {Escobar}, {Shenar}, \&
  {Torniamenti}}]{rastello+2023}
{Rastello}, S., {Iorio}, G., {Mapelli}, M., {et~al.} 2023, \mnras, 526, 740,
  \dodoi{10.1093/mnras/stad2757}

\bibitem[{{Rose} {et~al.}(2019){Rose}, {Naoz}, \& {Geller}}]{Ros+19}
{Rose}, S.~C., {Naoz}, S., \& {Geller}, A.~M. 2019, \mnras, 488, 2480,
  \dodoi{10.1093/mnras/stz1846}

\bibitem[{{Sahu} {et~al.}(2022){Sahu}, {Anderson}, {Casertano}, {Bond},
  {Udalski}, {Dominik}, {Calamida}, {Bellini}, {Brown}, {Rejkuba}, {Bajaj},
  {Kains}, {Ferguson}, {Fryer}, {Yock}, {Mr{\'o}z}, {Koz{\l}owski},
  {Pietrukowicz}, {Poleski}, {Skowron}, {Soszy{\'n}ski}, {Szyma{\'n}ski},
  {Ulaczyk}, {Wyrzykowski}, {Barry}, {Bennett}, {Bond}, {Hirao}, {Silva},
  {Kondo}, {Koshimoto}, {Ranc}, {Rattenbury}, {Sumi}, {Suzuki}, {Tristram},
  {Vandorou}, {Beaulieu}, {Marquette}, {Cole}, {Fouqu{\'e}}, {Hill}, {Dieters},
  {Coutures}, {Dominis-Prester}, {Bennett}, {Bachelet}, {Menzies}, {Albrow},
  {Pollard}, {Gould}, {Yee}, {Allen}, {Almeida}, {Christie}, {Drummond},
  {Gal-Yam}, {Gorbikov}, {Jablonski}, {Lee}, {Maoz}, {Manulis}, {McCormick},
  {Natusch}, {Pogge}, {Shvartzvald}, {J{\o}rgensen}, {Alsubai}, {Andersen},
  {Bozza}, {Novati}, {Burgdorf}, {Hinse}, {Hundertmark}, {Husser}, {Kerins},
  {Longa-Pe{\~n}a}, {Mancini}, {Penny}, {Rahvar}, {Ricci}, {Sajadian},
  {Skottfelt}, {Snodgrass}, {Southworth}, {Tregloan-Reed}, {Wambsganss},
  {Wertz}, {Tsapras}, {Street}, {Bramich}, {Horne}, {Steele}, \& {RoboNet
  Collaboration}}]{sahu+2022}
{Sahu}, K.~C., {Anderson}, J., {Casertano}, S., {et~al.} 2022, \apj, 933, 83,
  \dodoi{10.3847/1538-4357/ac739e}

\bibitem[{{Shahaf} {et~al.}(2023){Shahaf}, {Bashi}, {Mazeh}, {Faigler},
  {Arenou}, {El-Badry}, \& {Rix}}]{shahaf+2023}
{Shahaf}, S., {Bashi}, D., {Mazeh}, T., {et~al.} 2023, \mnras, 518, 2991,
  \dodoi{10.1093/mnras/stac3290}

\bibitem[{{Shariat} {et~al.}(2023){Shariat}, {Naoz}, {Hansen}, {Angelo},
  {Michaely}, \& {Stephan}}]{shariat+2023}
{Shariat}, C., {Naoz}, S., {Hansen}, B. M.~S., {et~al.} 2023, \apjl, 955, L14,
  \dodoi{10.3847/2041-8213/acf76b}

\bibitem[{{Shenar} {et~al.}(2022){Shenar}, {Sana}, {Mahy}, {El-Badry},
  {Marchant}, {Langer}, {Hawcroft}, {Fabry}, {Sen}, {Almeida}, {Abdul-Masih},
  {Bodensteiner}, {Crowther}, {Gieles}, {Gromadzki}, {H{\'e}nault-Brunet},
  {Herrero}, {de Koter}, {Iwanek}, {Koz{\l}owski}, {Lennon}, {Ma{\'\i}z
  Apell{\'a}niz}, {Mr{\'o}z}, {Moffat}, {Picco}, {Pietrukowicz}, {Poleski},
  {Rybicki}, {Schneider}, {Skowron}, {Skowron}, {Soszy{\'n}ski},
  {Szyma{\'n}ski}, {Toonen}, {Udalski}, {Ulaczyk}, {Vink}, \&
  {Wrona}}]{shenar+2022}
{Shenar}, T., {Sana}, H., {Mahy}, L., {et~al.} 2022, Nature Astronomy, 6, 1085,
  \dodoi{10.1038/s41550-022-01730-y}

\bibitem[{{Smith} \& {Conti}(2008)}]{smith&conti2008}
{Smith}, N., \& {Conti}, P.~S. 2008, \apj, 679, 1467, \dodoi{10.1086/586885}

\bibitem[{{Stegmann} {et~al.}(2022){Stegmann}, {Antonini}, \& {Moe}}]{Ste+22}
{Stegmann}, J., {Antonini}, F., \& {Moe}, M. 2022, \mnras, 516, 1406,
  \dodoi{10.1093/mnras/stac2192}

\bibitem[{{Tanikawa} {et~al.}(2024){Tanikawa}, {Cary}, {Shikauchi}, {Wang}, \&
  {Fujii}}]{tanikawa+2024}
{Tanikawa}, A., {Cary}, S., {Shikauchi}, M., {Wang}, L., \& {Fujii}, M.~S.
  2024, \mnras, 527, 4031, \dodoi{10.1093/mnras/stad3294}

\bibitem[{{The Astropy Collaboration} {et~al.}(2018){The Astropy
  Collaboration}, {Price-Whelan}, {Sip{\H o}cz}, {G{\"u}nther}, {Lim},
  {Crawford}, \& {Contributors}}]{astropy+2018}
{The Astropy Collaboration}, {Price-Whelan}, A.~M., {Sip{\H o}cz}, B.~M.,
  {et~al.} 2018, \aj, 156, 123, \dodoi{10.3847/1538-3881/aabc4f}

\bibitem[{{Toonen} {et~al.}(2022){Toonen}, {Boekholt}, \& {Portegies
  Zwart}}]{toonen+2022}
{Toonen}, S., {Boekholt}, T.~C.~N., \& {Portegies Zwart}, S. 2022, \aap, 661,
  A61, \dodoi{10.1051/0004-6361/202141991}

\bibitem[{{Toonen} {et~al.}(2020){Toonen}, {Portegies Zwart}, {Hamers}, \&
  {Bandopadhyay}}]{toonen+2020}
{Toonen}, S., {Portegies Zwart}, S., {Hamers}, A.~S., \& {Bandopadhyay}, D.
  2020, \aap, 640, A16, \dodoi{10.1051/0004-6361/201936835}

\bibitem[{Virtanen {et~al.}(2020)Virtanen, Gommers, Oliphant, Haberland, Reddy,
  Cournapeau, Burovski, Peterson, Weckesser, Bright, Walt, Brett, Wilson, K,
  Mayorov, Nelson, Jones, Kern, Larson, C, Polat, Feng, Moore, VanderPlas,
  Laxalde, Perktold, Cimrman, Henriksen, E, Harris, Archibald, Ribeiro,
  Pedregosa, Mulbregt, Vijaykumar, Bardelli, Rothberg, Hilboll, Kloeckner,
  Scopatz, Lee, Rokem, C, Fulton, Masson, Häggström, Fitzgerald, Nicholson,
  Hagen, Pasechnik, Olivetti, Martin, Wieser, Silva, Lenders, Wilhelm, G,
  Price, Ingold, Allen, Lee, Audren, Probst, Dietrich, Silterra, Webber,
  Slavič, Nothman, Buchner, Kulick, Schönberger, Cardoso, Reimer, Harrington,
  Rodríguez, Nunez-Iglesias, Kuczynski, Tritz, Thoma, Newville, Kümmerer,
  Bolingbroke, Tartre, Pak, Smith, Nowaczyk, Shebanov, Pavlyk, Brodtkorb, Lee,
  McGibbon, Feldbauer, Lewis, Tygier, Sievert, Vigna, Peterson, More, Pudlik,
  Oshima, Pingel, Robitaille, Spura, Jones, Cera, Leslie, Zito, Krauss,
  Upadhyay, Halchenko, \& Vázquez-Baeza}]{2020SciPy-NMeth}
Virtanen, P., Gommers, R., Oliphant, T.~E., {et~al.} 2020, Nature Methods, 17,
  261, \dodoi{10.1038/s41592-019-0686-2}

\end{thebibliography}


\end{document}